\documentclass[12pt,a4paper]{article}
\usepackage{amssymb, amsmath, graphicx}
\usepackage{float}

\def\eslt{E_T^{\rm miss}}

\def\to{\rightarrow}

\def\bi{\begin{itemize}}
 \def\ei{\end{itemize}}
\def\te{\tilde e}

\def\c1p{C1^\prime}
\def\msq3{\overline{m}_{\tilde{q}}(3)}
\def\ta{\tilde a}

\def\tu{\tilde u}

\def\ta{\tilde a}

\def\tb{\tilde b}

\def\tst{\tilde t}
\def\ttau{\tilde \tau}

\def\tg{\tilde g}

\def\tq{\tilde q}

\def\tw{\widetilde W}
\def\tz{\widetilde Z}
\def\alt{\lesssim}
\def\agt{\gtrsim}
\def\be{\begin{equation}}  
\def\ee{\end{equation}}  
\def\bea{\begin{eqnarray}}  
\def\eea{\end{eqnarray}}  

\def\sps1ap{SPS1a$^\prime$}

\def\Ryuk{R_{\rm yuk}}
\def\bsg{{\rm BR}(B\to X_s\gamma)}
\def\bmm{{\rm BR}(B_s\to\mu^+\mu^-)}

\newcommand\prd[3]{{Phys.\ Rev.\ }{\bf D#1} (#2) #3}

\newcommand\prl[3]{{Phys.\ Rev.\ Lett.\ }{\bf #1} (#2) #3}
\newcommand\plb[3]{{Phys.\ Lett.\ }{\bf B#1} (#2) #3}

\newcommand\jhep[3]{{JHEP}{\bf #1} (#2) #3}
\newcommand\jpg[3]{{J.\ Phys.\ G}{\bf #1} (#2) #3}

\newcommand\npb[3]{{Nucl.\ Phys.\ }{\bf B#1} (#2) #3}

\newcommand{\hepph}[1]{hep-ph/#1}

\newcommand\arXivid[1]{arXiv:#1}

\begin{document}

\begin{titlepage}

\begin{center}

\vspace*{4mm}

{\LARGE\bf Yukawa-unified natural supersymmetry}\\

\vspace*{12mm} \renewcommand{\thefootnote}{\fnsymbol{footnote}}
{\large Howard Baer$^{1}$\footnote[1]{Email: baer@nhn.ou.edu },
Sabine Kraml$^{2}$\footnote[2]{Email: sabine.kraml@lpsc.in2p3.fr}, 
Suchita Kulkarni$^{2}$\footnote[3]{Email: suchita.kulkarni@lpsc.in2p3.fr} } \\
\vspace{0.8cm} \renewcommand{\thefootnote}{\arabic{footnote}}
{\it 
$^1$Dept. of Physics and Astronomy,
University of Oklahoma, Norman, OK 73019, USA \\[2mm]
$^2$Laboratoire de Physique Subatomique et de Cosmologie, UJF Grenoble 1,
CNRS/IN2P3, INPG, 53 Avenue des Martyrs, F-38026 Grenoble, France
}

\end{center}

\vspace{0.6cm}

\begin{abstract}
\noindent 
Previous work on $t-b-\tau$ Yukawa-unified supersymmetry, as expected 
from SUSY GUT theories based on the gauge group $SO(10)$, 
tended to have exceedingly large electroweak fine-tuning (EWFT).
Here, we examine supersymmetric models where we simultaneously require 
low EWFT (``natural SUSY'') and a high degree of Yukawa coupling unification, 
along with a light Higgs scalar with $m_h\sim 125$ GeV. 
As Yukawa unification requires large $\tan\beta\sim 50$, while EWFT requires rather light
third generation squarks and low $\mu\approx100-250$~GeV, 
$B$-physics constraints from $\bsg$ and $\bmm$ can be severe.
We are able to find models with EWFT $\Delta\lesssim 50-100$ (better than 1--2\% EWFT) and
with Yukawa unification as low as $\Ryuk\sim 1.3$ (30\% unification) if $B$-physics constraints are imposed. 
This may be improved to $\Ryuk\sim 1.2$ if additional small flavor violating terms conspire
to improve accord with $B$-constraints. 
We present several Yukawa-unified natural SUSY (YUNS) benchmark points.
LHC searches will be able to access gluinos in the lower $1-2$ TeV portion of their predicted mass range 
although much of YUNS parameter space may lie beyond LHC14 reach. 
If heavy Higgs bosons can be accessed at a high rate, then the rare $H,\, A\to\mu^+\mu^-$ decay
might allow a determination of $\tan\beta\sim 50$ as predicted by YUNS models.
Finally, the predicted light higgsinos should be accessible to a linear $e^+e^-$ 
collider with $\sqrt{s}\sim 0.5$ TeV. 

\vspace{0.8cm}


\end{abstract}

\end{titlepage}

\section{Introduction}
\label{sec:intro}

A striking feature in nature is that all the fermions of each generation fill out a complete
16-dimensional spinor multiplet of the gauge group $SO(10)$~\cite{so10}. 
While ordinary grand unified theories (GUTs) suffer from the notorious gauge hierarchy problem, 
supersymmetric (SUSY) GUTs not only tame this hierarchy problem~\cite{witten}, 
but they also receive support from the well-known unification of gauge couplings~\cite{gauge}.
In the simplest $SO(10)$ SUSY GUT theories, where both MSSM Higgs doublets $H_u$ and $H_d$ occupy
the same 10-dimensional representation, one also expects unification of third generation 
Yukawa couplings $f_t$, $f_b$ and $f_\tau$ at $M_{\rm GUT}\simeq 2\times 10^{16}$ GeV~\cite{old,new}. 
The $t-b-\tau$ Yukawa coupling unification is highly sensitive to both 2-loop renormalization group
running (RGEs) and to threshold corrections when transitioning between MSSM and SM effective theories
at the SUSY particle mass scale. Thus, the entire SUSY mass spectrum enters into a precise computation of
Yukawa coupling unification.

Many groups have explored $t-b-\tau$ Yukawa unification (YU) in SUSY 
theories~\cite{hb,bf,abbbft,bdr,bkss,alt,bartol,hall}. 
It has been found that, for $\mu>0$, YU can occur at the few percent level in either the Higgs splitting (HS) model or the DR3 model (D-term splitting, right-hand neutrino effects and third generation splitting),
provided that the GUT scale soft SUSY breaking (SSB) terms are related as
\be
  A_0^2\simeq 2m_{10}^2\simeq 4m_{16}^2 , 
  \label{eq:YUparamRel}
\ee
with $A_0<0$. Moreover, the GUT scale Higgs splitting $m_{H_u}^2<m_{H_d}^2$ is needed to allow for an appropriate
radiative breakdown of electroweak symmetry. In these models, first generation squarks and sleptons
are required to be in the multi-TeV range~\cite{abbbft,bkss} while 
third generation sfermions are driven to much lighter TeV-scale masses. 
A benefit of these models is that the light SUSY Higgs boson mass $m_h$ tends naturally to be in the 
125 GeV range,\footnote{The large negative $A_0$ in Eq.~(\ref{eq:YUparamRel}) leads to maximal stop mixing, 
see {\it e.g.}~\cite{h125,Brummer:2012ns}, thus increasing $m_h$ to the desired range.}
as required by the recent LHC discovery~\cite{atlashiggs,cmshiggs}.
The gaugino masses from YU SUSY are expected to be quite light, with
typically $m_{\tg}\alt 500$ GeV (now excluded by LHC searches for gluino pair production along with
cascade decay into states containing $b$-quarks\cite{ATLAS:2012ah}), although solutions with heavier gluinos $\sim 1-2$ TeV 
can also be found~\cite{Baer:2012cp}. In all these cases, the superpotential $\mu$ parameter, which is extracted
from the electroweak minimization conditions, occupies values in the 1--10 TeV regime, leading to 
severe electroweak fine-tuning (EWFT).

In this work, we examine to what extent it is possible to reconcile YU with low EWFT. 
Minimization of the SUSY scalar potential allows one to relate the $Z$ mass scale to the 
superpartner mass scale via the well-known relation
\be
\frac{1}{2} M_Z^2 = \frac{(m^2_{H_d}+\Sigma_d)-(m^2_{H_u}+\Sigma_u)\tan^2\beta}{(\tan^2\beta-1)} - \mu^2 .
\label{eq:ewsb}
\ee
The radiative corrections $\Sigma_u$ and $\Sigma_d$ are given in the 1-loop 
approximation of the Higgs effective potential by: 
\be 
\Sigma_{u,d} = \frac{1}{v_{u,d}}\frac{\partial\Delta V}{\partial H_{u,d}},
\label{eq:sigma}
\ee
where $\Delta V$ is the one-loop correction to the tree-level potential, and the derivative is 
evaluated in the physical vacuum: {\it i.e.} the fields are set to their vacuum expectation 
values after evaluating the derivative. 
At the one-loop level and in the limit of setting first/second generation Yukawa
couplings to zero, $\Sigma_u$
contains 18 and $\Sigma_d$ contains 19 separate contributions from
various particles/sparticles~\cite{Baer:2012up}.
We include contributions from $W^\pm$, $Z$, $\tst_{1,2}$, $\tb_{1,2}$, $\ttau_{\
1,2}$, $\tw_{1,2}$,
$\tz_{1,2,3,4}$, $t$, $b$ and $\tau$, $h$, $H$ and $H^\pm$. We adopt a scale choice
$Q^2=m_{\tst_1}m_{\tst_2}$ to minimize the largest of the logarithms.
The dominant contribution to the terms $\Sigma_{u,d}$ arise from superpotential Yukawa interactions of 
third generation squarks involving the top quark Yukawa coupling. 
For instance, the dominant contribution to $\Sigma_u$
is given by
\be
\Sigma_u (\tst_{1,2} )=\frac{3}{16\pi^2}F(m_{\tst_{1,2}}^2)
\left[ f_t^2-g_Z^2\mp \frac{f_t^2 A_t^2-8g_Z^2(\frac{1}{4}-\frac{2}{3}x_W)\Delta_t}{m_{\tst_2}^2-m_{\tst_1}^2} \right] \,,
\label{eq:Sigu}
\ee
where $\Delta_t=(m_{\tst_L}^2-m_{\tst_R}^2)/2+m_Z^2\cos                         
2\beta(\frac{1}{4}-\frac{2}{3}x_W)$, $g_Z^2=(g^2+g^{\prime 2})/8$ and
$x_W\equiv \sin^2\theta_W$ and $F(m^2)=m^2(\log \frac{m^2}{Q^2}-1)$.
This expression thus grows quadratically with the stop mass. 

We adopt the fine-tuning measure from~\cite{Baer:2012up}, which requires that each of the 
40 terms on the right-hand-side (RHS) of Eq.~(\ref{eq:ewsb}) should be of order $\sim m_Z^2/2$.
Labeling each term as $C_i$ (with $i=H_d,\ H_u,\ \mu ,\                         
\Sigma_d^d(\tst_1),\ \Sigma_u^u(\tst_1),\ etc.$), we may require
$C_{max}\equiv max|C_i|<\Lambda_{max}^2$, where $\Lambda_{max}\sim              
100-300$ GeV, depending on how much EWFT one is willing to
tolerate. 
This measure of fine-tuning is similar to (but not exactly the
same as) Kitano--Nomura~\cite{Kitano:2005wc} but different from
Barbieri--Giudice~\cite{Barbieri:1987fn} beyond the tree-level.  
In the following, we will use the fine-tuning parameter  
\be
 \Delta =C_{max}/(m_Z^2/2),
\ee 
where lower values of $\Delta$ correspond
to less fine-tuning, and {\it e.g.}\ $\Delta =20$ would correspond to
$\Delta^{-1}=5\%$ fine-tuning. 

Our goal in this paper is to search for parameter choices which
\begin{enumerate}
\item  Maximize the degree of Yukawa coupling unification, i.e.\ minimize 
\be
\Ryuk=\frac{{\rm max}(f_t,\, f_b,\, f_\tau )}{{\rm min}(f_t,\, f_b,\, f_\tau )}
\ee
with each Yukawa coupling evaluated at the GUT scale. Thus, a value of $\Ryuk=1$ would give 
perfect Yukawa coupling unification. 
\item Have as low EWFT as possible (in practice, we will require $\Delta\lesssim 100$, 
or better than 1\% EWFT). 
\item Have $m_h\approx 125$ GeV in accord with the recent LHC discovery of a Higgs-like resonance.
In practice, we will require $122\ {\rm GeV}<m_h<128$ GeV to allow for a roughly 2--3 GeV
error in the RG-improved one-loop effective potential calculation of the Higgs mass $m_h$.
\end{enumerate}

We recognize that in addition to EWFT, there also exists a fine-tuning associated with
generating particular weak scale SUSY spectra from distinct GUT scale parameters~\cite{Barbieri:1987fn}, 
(see also~\cite{Ghilencea:2012gz,Wymant:2012zp} for related discussions).
Here we adopt the less restrictive {\it weak scale} fine-tuning condition, which nonetheless turns out to be
indeed very restrictive. In this vein, we regard particular GUT scale parameters as merely a parametrization
of our ignorance of the mechanism of SUSY breaking and soft term generation.\footnote{The relation between GUT scale fine-tuning and the $b$--$\tau$ Yukawa coupling ratio was studied in \cite{Antusch:2011xz}.}
For instance, in this paper we will require rather low values of superpotential $\mu$ parameter to avoid excessive EWFT
in Eq.~(\ref{eq:ewsb}). In generic SUSY models, the value of $\mu$ is expected to be of order
$M_{\rm Planck}$ since it is a dimensionful SUSY-preserving parameter. Excessively large $\mu$ can be avoided
ala Giudice-Masiero~\cite{Giudice:1988yz} where the $\mu$ superpotential term is forbidden by some high scale
symmetry, but then is regenerated as a soft SUSY breaking term via a Higgs--Higgs coupling to 
the hidden sector.
    
For the remainder of this paper, in Section~\ref{sec:scan} we present details of our scan over
SUSY parameter space, and which constraints are invoked in our analysis. 
In Section~\ref{sec:results}, we present the results of our parameter space scans. 
We will find that requiring $\Delta\alt 100$ and $m_h\sim 125$ GeV only allows for
$\Ryuk$ as low as $\sim 1.2-1.3$. While this degree of Yukawa unification is not optimal, 
we feel it is still useful in that it might guide model builders towards models including
additional GUT scale threshold corrections or extra matter or above-GUT-scale running
which may ameliorate the situation. 
In Section~\ref{sec:exp}, we discuss observable consequences of YUNS for LHC, ILC and dark matter
searches. 
We pay some attention to methods which might allow one to distinguish YUNS from generic NS models 
at lower $\tan\beta $ values.
In Section~\ref{sec:conclude} we present a summary and conclusions.

\section{Parameter space and Yukawa unification}
\label{sec:scan}

For our calculations, we adopt the ISAJET\,7.83~\cite{isajet} SUSY
spectrum generator ISASUGRA~\cite{isasugra}. ISASUGRA begins the
calculation of the sparticle mass spectrum with input $\overline{DR}$
gauge couplings and $f_b$, $f_\tau$ Yukawa couplings at the scale
$Q=M_Z$ ($f_t$ running begins at $Q=m_t$) and evolves the 6 couplings up
in energy to scale $Q=M_{\rm GUT}$ (defined as the value $Q$ where
$g_1=g_2$) using two-loop RGEs.  
At $Q=M_{\rm GUT}$, we input the soft SUSY breaking parameters as boundary
conditions, and evolve the set of 26 coupled MSSM
RGEs~\cite{mv} back down in scale to $Q=M_Z$.  Full two-loop
MSSM RGEs are used for soft term evolution, while the gauge and Yukawa
coupling evolution includes threshold effects in the one-loop
beta-functions, so the gauge and Yukawa couplings transition smoothly
from the MSSM to SM effective theories as different mass thresholds are
passed.  In ISASUGRA, the values of SSB terms of sparticles which mix are frozen out
at the scale $Q\equiv M_{SUSY}=\sqrt{m_{\tst_L} m_{\tst_R}}$, while
non-mixing SSB terms are frozen out at their own mass
scale~\cite{isasugra}.  The scalar potential is minimized using the
RG-improved one-loop MSSM effective potential evaluated at an optimized
scale $Q=M_{SUSY}$ which accounts for leading two-loop
effects~\cite{haber}.  Once the tree-level sparticle mass spectrum is
computed, full one-loop radiative corrections are calculated for all
sparticle and Higgs boson masses, including complete one-loop weak scale
threshold corrections for the top, bottom and tau masses at scale
$Q=M_{SUSY}$~\cite{pbmz}.  Since the GUT scale Yukawa couplings are
modified by the threshold corrections, the ISAJET RGE solution must be 
imposed iteratively with successive up-down running until a convergent
sparticle mass solution is found.  Since ISASUGRA uses a ``tower of             
effective theories'' approach to RG evolution, we expect a more accurate
evaluation of the sparticle mass spectrum for models with split spectra
(this procedure sums the logarithms of potentially large ratios of 
sparticle masses) than with programs which make an all-at-once transition
from the MSSM to SM effective theories.
The fine-tuning measure $\Delta$ described in Sec.~\ref{sec:intro} has been implemented in 
ISAJET\,7.83~\cite{isajet}.

In models of ``natural SUSY'' (NS)~\cite{ccn,Kitano:2005wc,Baer:2010ny, Asano:2010ut,bbh,NS,Baer:2012uy,Baer:2012up}, 
the first requirement to gain a low EWFT is that the 
$\mu$ parameter be of the order of $\sim M_Z$, while in Yukawa-unified SUSY, it is necessary
to invoke some manner of Higgs soft term splitting at the GUT scale in order to obtain
raditive EWSB. For these reasons, the model parameter space chosen in this study is the two-parameter
non-universal Higgs model (NUHM2) where the weak scale values of $\mu$ and $m_A$ are input {\it in lieu}
of the GUT scale values of $m_{H_u}^2$ and $m_{H_d}^2$. In addition, to allow for (sub)TeV-scale third generation 
masses (as required by EWFT from Eq.~(\ref{eq:ewsb}) along with at least a partial decoupling 
solution to the SUSY flavor and CP problems, we allow for split first/second and third generations 
at the GUT scale. Thus, the parameter space we choose is given by 
\begin{equation}
   m_{16}(1,2),\ m_{16}(3),\ m_{1/2},\ A_0,\ \tan\beta,\ \mu,\ m_A \;.
\end{equation}
Here, $m_{16}(1,2)$ and $m_{16}(3)$ are the first/second and third generation sfermion soft masses, respectively; 
$m_{1/2}\equiv M_1=M_2=M_3$ is the universal gaugino mass parameter;  and 
$A_0\equiv A_t=A_b=A_\tau$ is the universal trilinear coupling. These parameters are defined at $M_{\rm GUT}$, 
while $\tan\beta$, $\mu$, and $m_A$ are defined at the weak scale. 
The top quark mass is set to $m_t=173.2$ GeV.


We search for mass spectra with low EWFT $\Delta$ and low $\Ryuk$ by performing a vast random scan over the following parameter ranges (masses in GeV units):
\bea
&  5000\ {\rm GeV} < m_{16}(1,2) < 20000\ {\rm GeV}, &  \nonumber\\
&  0 <  m_{16}(3) < 20000\ {\rm GeV}, &  \nonumber\\
& 300\ {\rm GeV} < m_{1/2} < 2000\ {\rm GeV}, &  \nonumber\\
& -2 < A_0/m_{16}(3) < 1, &  \\
&  50 < \tan\beta < 60, &  \nonumber\\
&  100\ {\rm GeV} < \mu < 250\ {\rm GeV}, &  \nonumber\\
&  500\ {\rm GeV} < m_A < 5000\ {\rm GeV}. \nonumber
  \label{eq:ScanRanges}
\eea
The lower limit on $m_{1/2}$ comes from the approximate LHC bound of 
$m_{\tg}\agt 900$~GeV (for $m_{\tq}\gg m_{\tg}$)~\cite{lhc_mgl}, 
while the lower bound on $m_A$ comes from LHC searches for $A,\,H\to\tau^+\tau^-$ which
require $m_A>500$~GeV at $\tan\beta\sim 50$~\cite{Chatrchyan:2012vp}.
We require of our solutions that 
\begin{enumerate}
\item electroweak symmetry be radiatively broken (REWSB), 
\item the neutralino $\tz_1$ is the lightest MSSM particle, 
\item the light chargino mass obeys the rather model independent LEP2 limit that 
$m_{\tw_1}>103.5$~GeV~\cite{lep2ino}, and
\item the light Higgs mass falls within the window $m_h=122\mbox{--}128$~GeV, where we adopt 
$\pm 3$~GeV as theoretical error on the Higgs mass calculation.
\end{enumerate}

Regarding $B$-physics constraints, we consider  
$\bsg = (3.55\pm 0.34) \times 10^{-4}$,  where experimental and theoretical uncertainties have been added in squares, 
and $\bmm<4.2 \times 10^{ -9}$ at 95\% CL. 
In the following we will impose the $\bsg$ constraint only at the $3\sigma$ level (the reason for this is that calculating $\bsg$ with both IsaTools and SuperISO, we observe deviations of the order of 5\% in the relevant region of parameter space).
For the $\bmm$ constraint, we assume a  theoretical uncertainty of 20\%. 
This leads to the the following limits
\bea
  & \bsg = [2.53,\ 4.57] \times 10^{-4} \,, & \nonumber \\
  &\bmm < 5.04 \times 10^{ -9} \,, & 
  \label{eq:flavor}
\eea
which we will use throughout the numerical analysis. 

Regarding neutralino relic density, we remark here that models of natural SUSY contain a
higgsino-like lightest neutralino with thermal abundance of typically 
$\Omega_{\tz_1}^{\rm TP}h^2\sim {\cal O}(10^{-3}-10^{-2})$.
This thermal under-abundance can be regarded as a positive feature of NS models in the sense that
if one invokes the axion solution to the strong CP problem, then one expects mixed axion-higgsino
dark matter, where the higgsino portion is typically enhanced by thermal axino production and decay
to higgsinos in the early universe. Thus, a thermal under-abundance leaves room for additional non-thermal
higgsino production plus an axion component to the dark matter~\cite{Baer:2011uz}. 

\section{Scan results}  
\label{sec:results}

\begin{figure}[t]\centering
\includegraphics[width=0.6\textwidth]{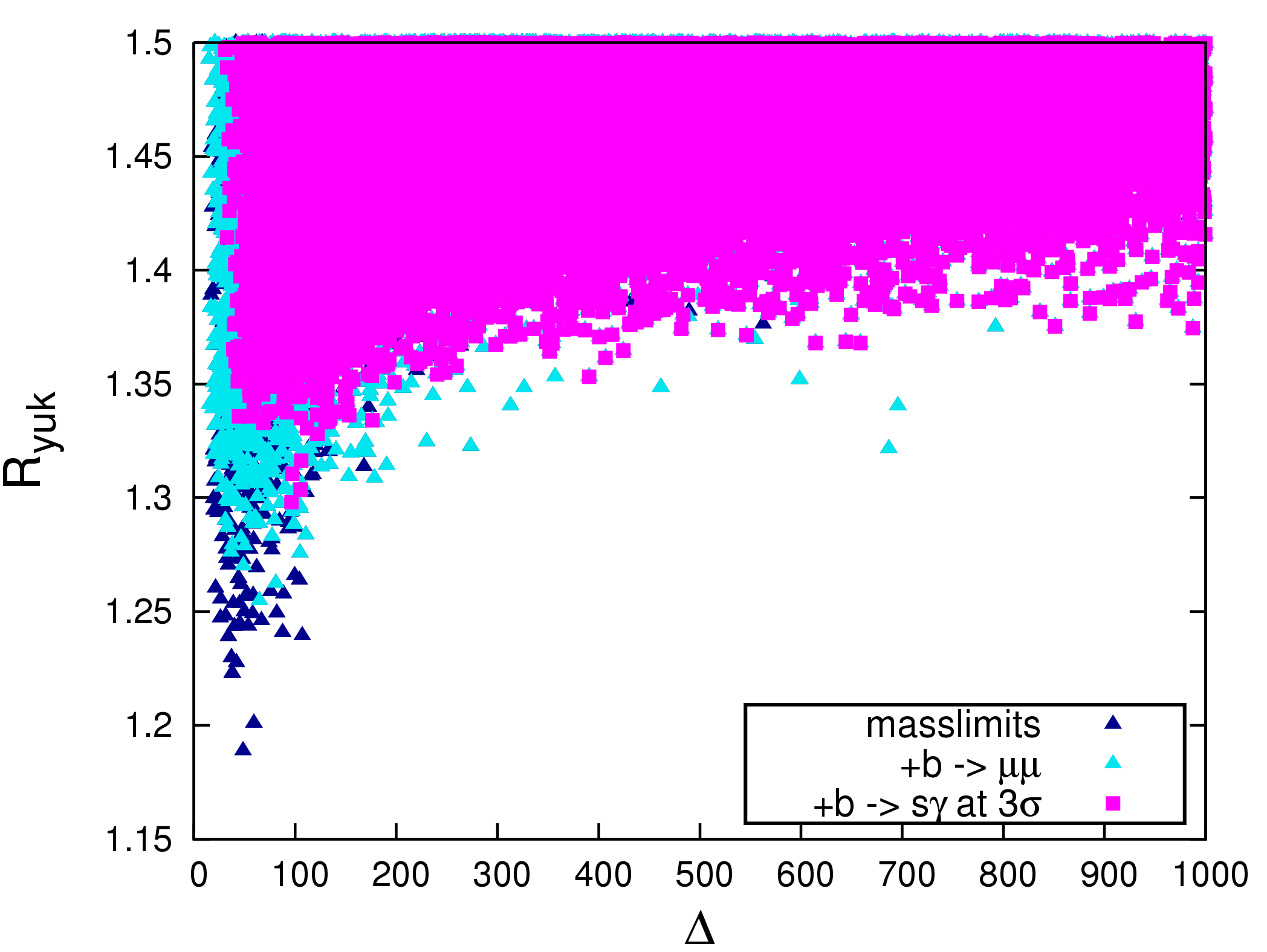}  
\caption{Scatter plot of $\Ryuk$ versus $\Delta$ from the scan defined in Eq.~(2.2) 
The dark blue triangles violate the $B$-physics constraints of Eq.~(2.3) 
The light blue triangles obey the $\bmm$ constraint, 
but deviate from the measured $\bsg$ by more than $3\sigma$. 
Finally, the pink squares satisfy both the $\bmm$ and $\bsg$ constraints. 
Only points with $\Delta<1000$ and $R_{yuk} <1.5$ are shown. 
\label{fig:RyukDelta} }
\end{figure}

As our first result, we show in Fig.~\ref{fig:RyukDelta} points from our parameter space scan in the 
$\Ryuk\ vs.\ \Delta$ plane. 
All points have $m_h=122\mbox{--}128$~GeV and obey the current LEP and LHC SUSY mass limits. 
The dark blue triangles however violate the $B$-physics constraints of Eq.~(\ref{eq:flavor}). 
These points may still be valid if additional small flavor violating terms are allowed in the theory.
The light blue triangles obey the $\bmm$ constraint, but deviate from the measured $\bsg$ by more than $3\sigma$. 
Finally, the pink squares satisfy both the $\bmm$ and $\bsg$ constraints. 
This color scheme is used throughout the remainder of the paper. 
We see already from this plot that flavor physics constraints significantly affect the parameter space of 
Yukawa-unified natural SUSY. This is to be expected, since naturalness requires lighter third generation 
squarks while Yukawa unification requires $\tan\beta\sim 50$: both these effects bolster SUSY contributions
to $B$-physics observables. (Analogous observations were made in \cite{bkss,Albrecht:2007ii} in the context of generic YU models.)
Without flavor constrains, we can obtain $\Ryuk$ as low as $\sim 1.2$. 
The $\bmm$ constraint pushes this up to $\Ryuk\gtrsim 1.27$, and the $\bsg$ constraint to $\Ryuk\gtrsim 1.3$. 
Aside from the $B$-physics constraints, it is intriguing that points with lowest $\Delta$ also have 
lowest values of $\Ryuk$. This is because low $\Delta$ requires light third generation squark masses, 
while at the same time Yukawa unification requires large SUSY threshold corrections 
which also require lighter third generation squarks.
    
\begin{figure}[t]\centering
  \includegraphics[width=0.37\textwidth,angle=-90]{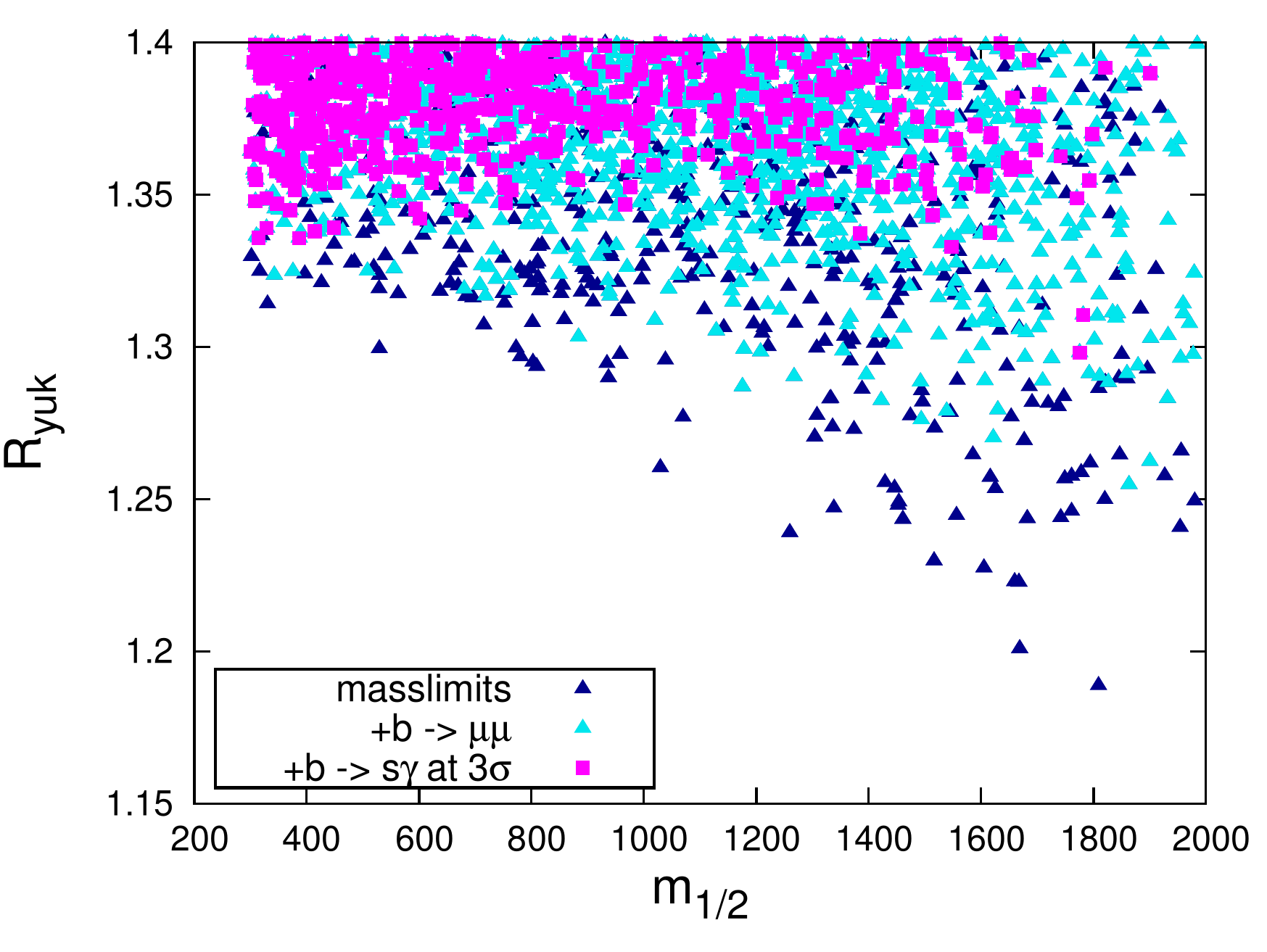}  
  \includegraphics[width=0.37\textwidth,angle=-90]{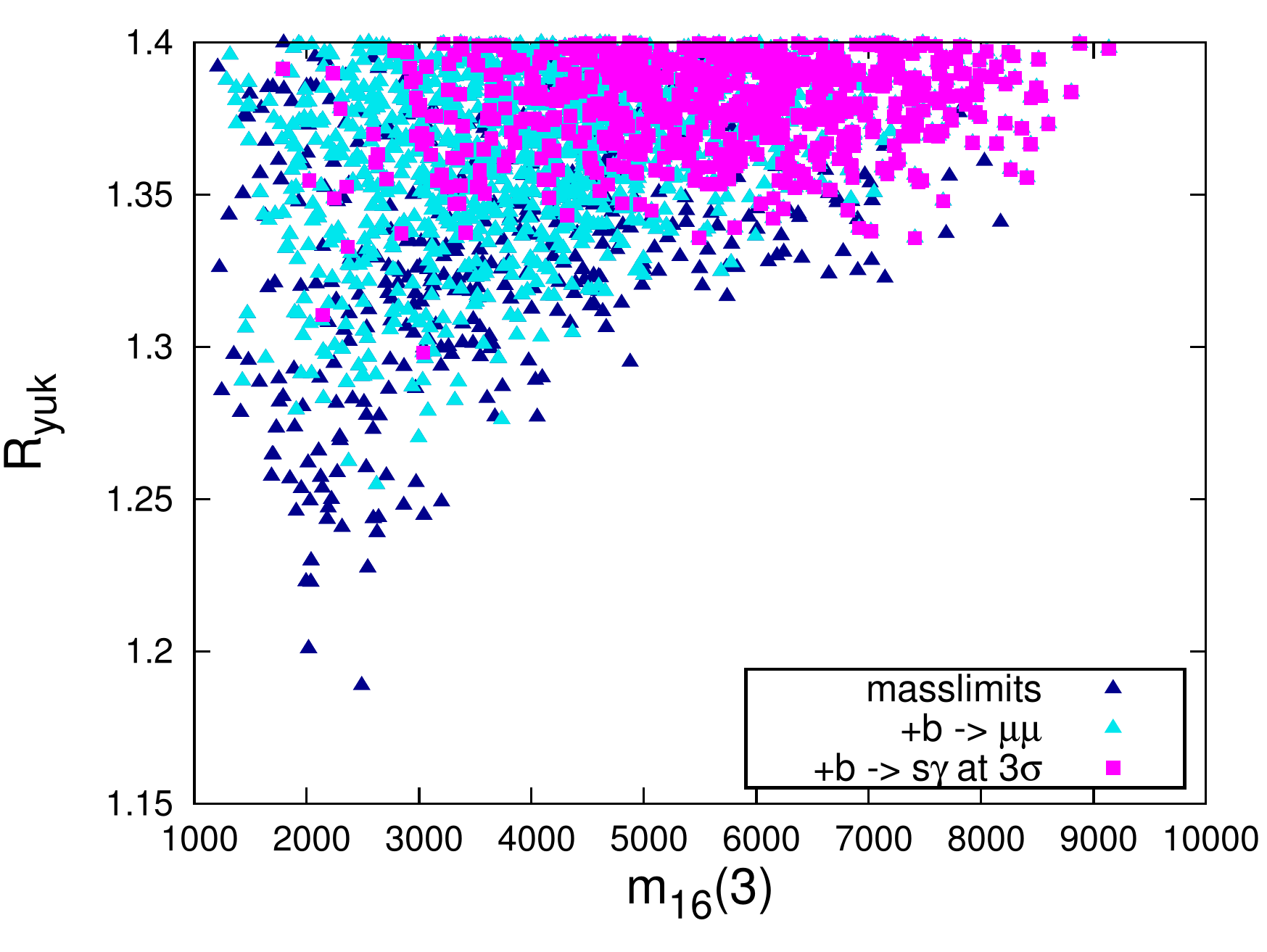}  
  \includegraphics[width=0.37\textwidth,angle=-90]{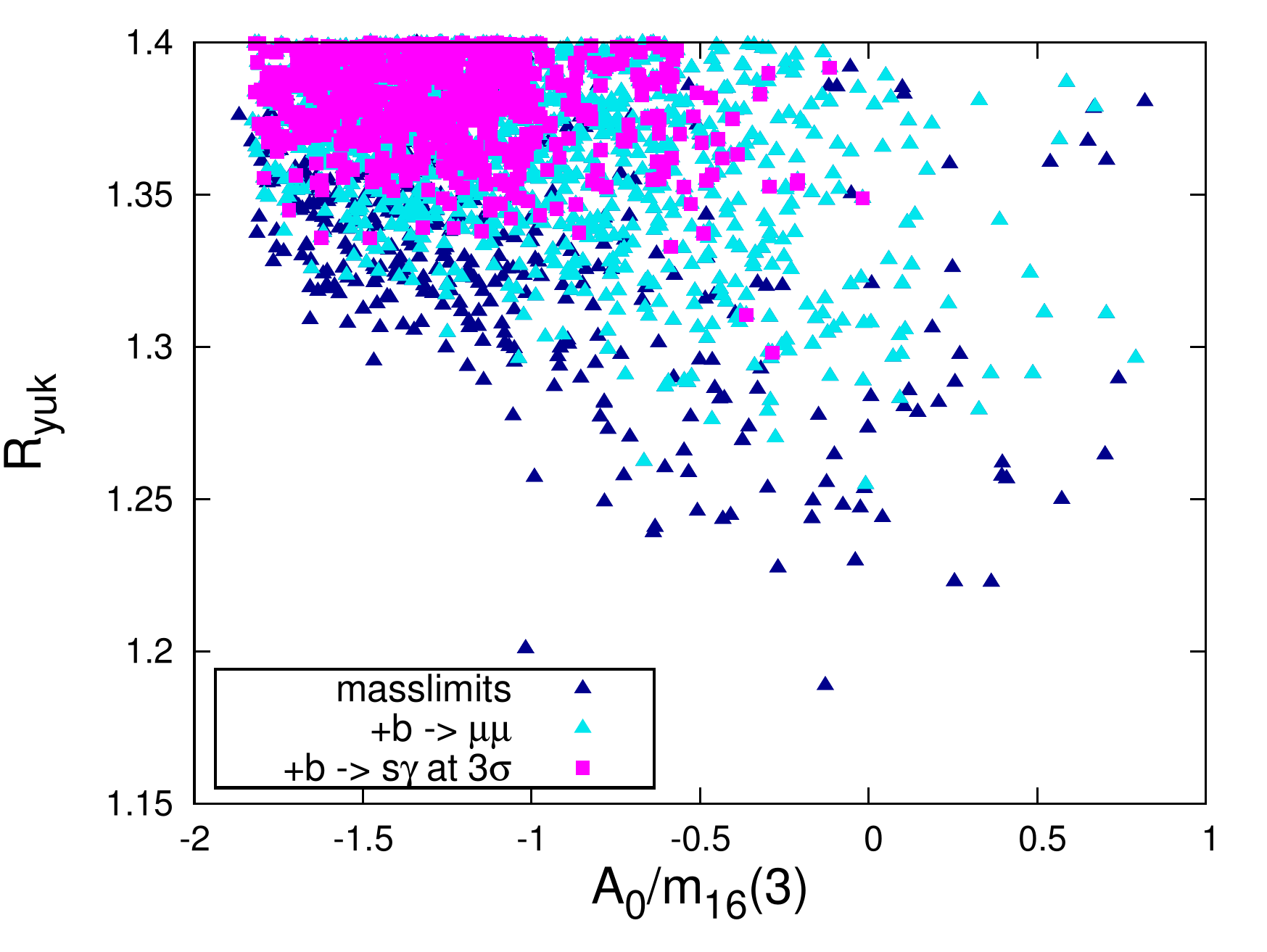}  
  \includegraphics[width=0.37\textwidth,angle=-90]{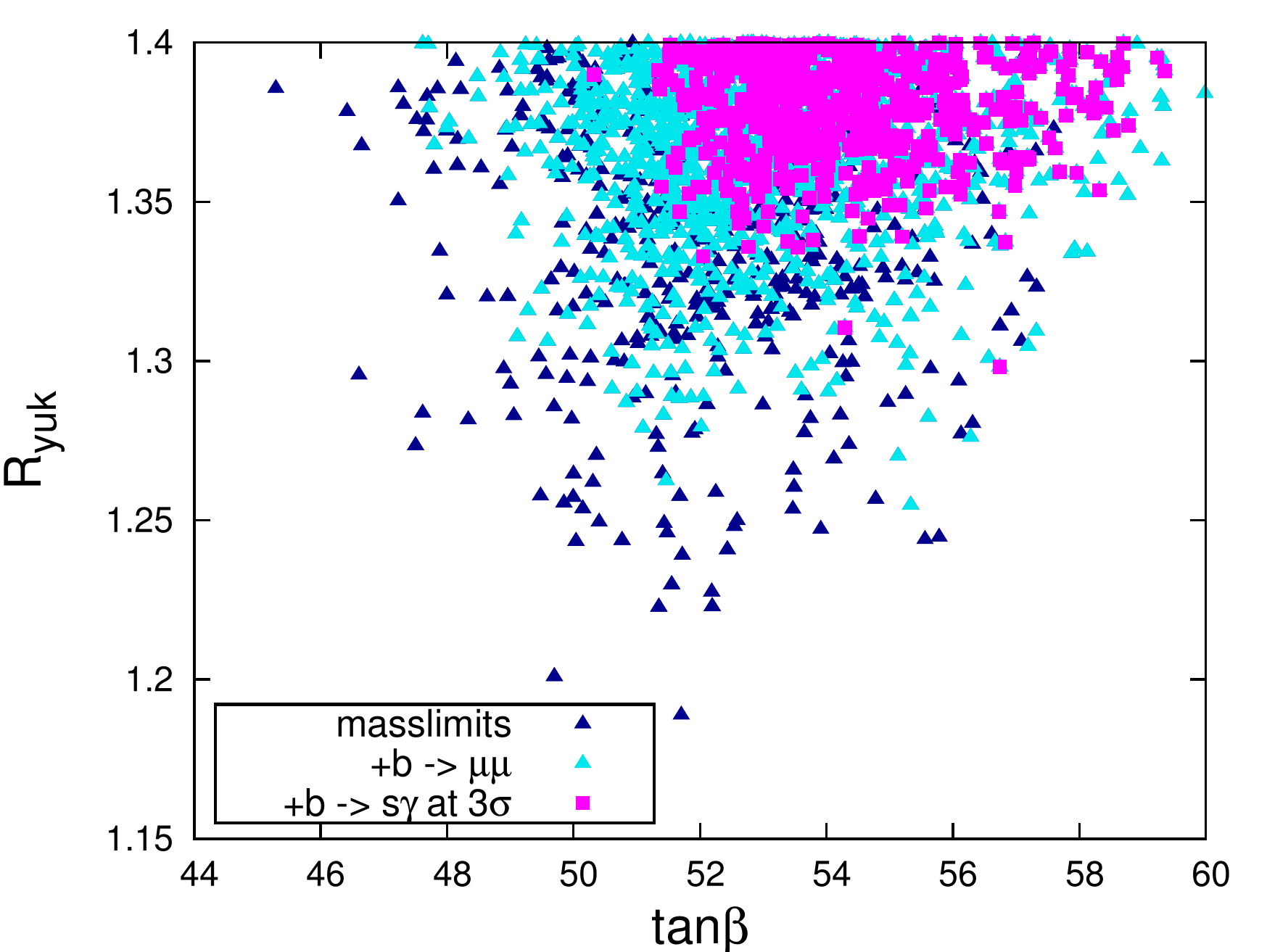}  
\caption{Dependence of $\Ryuk$ on $m_{1/2}$, $m_{16}(3)$, $A_0/m_{16}(3)$ and $\tan\beta$, for $\Delta<100$. 
Same color code as in Fig.~\ref{fig:RyukDelta}. \label{fig:RyukInpar} }
\end{figure}
  
In Fig.~\ref{fig:RyukInpar}, we show the value of $\Ryuk$ versus various input parameters for $\Delta<100$ 
(less than 1\% fine-tuning). Here, the behavior deviates considerably from $t-b-\tau$ unified models
with large $\mu>0$ and a spectrum derived from the radiatively driven inverted scalar mass 
hierarchy~\cite{bf,abbbft,bkss}. 
From frame {\it a}), we see that YUNS models actually prefer large $m_{1/2}$ whereas generic 
Yukawa unified SUSY models (with $\Ryuk\alt 1.1$ but arbitrary EWFT), YUS for short, prefer low $m_{1/2}$.
The preference for large $m_{1/2}$ helps to avoid LHC constraints on the gluino mass. If we impose the
$B$-physics constraints, then the distribution flattens out with some preference for lower $m_{1/2}$ values.
In frame {\it b}), where $\Ryuk$ is plotted {\it vs.} $m_{16}(3)$, we see that lowest $\Ryuk$
values prefer $m_{16}(3)\sim 2$~TeV, which leads to rather light third generation squarks and typically
violation of $B$-physics constraints. If we respect $B$-constraints, then larger values of $m_{16}(3)\sim 3-7$ TeV
are preferred, at the cost of larger values of $\Ryuk$.
In frame {\it c}), we plot versus $A_0/m_{16}(3)$. For YUS models, there is a strong  preference 
for $A_0\sim -2m_{16}$~\cite{bf}, while for YUNS, the lowest $\Ryuk$ values are obtained for smaller $|A_0|$. Imposing $B$-constraints, the preference moves to $A_0\sim (-2\ {\rm to}\ 0)\times m_{16}(3)$.
Frame {\it d}) shows $\Ryuk\ vs.\ \tan\beta$. As expected, $\tan\beta\sim 50$ is preferred.

\begin{figure}[t]\centering
  \includegraphics[width=0.37\textwidth,angle=-90]{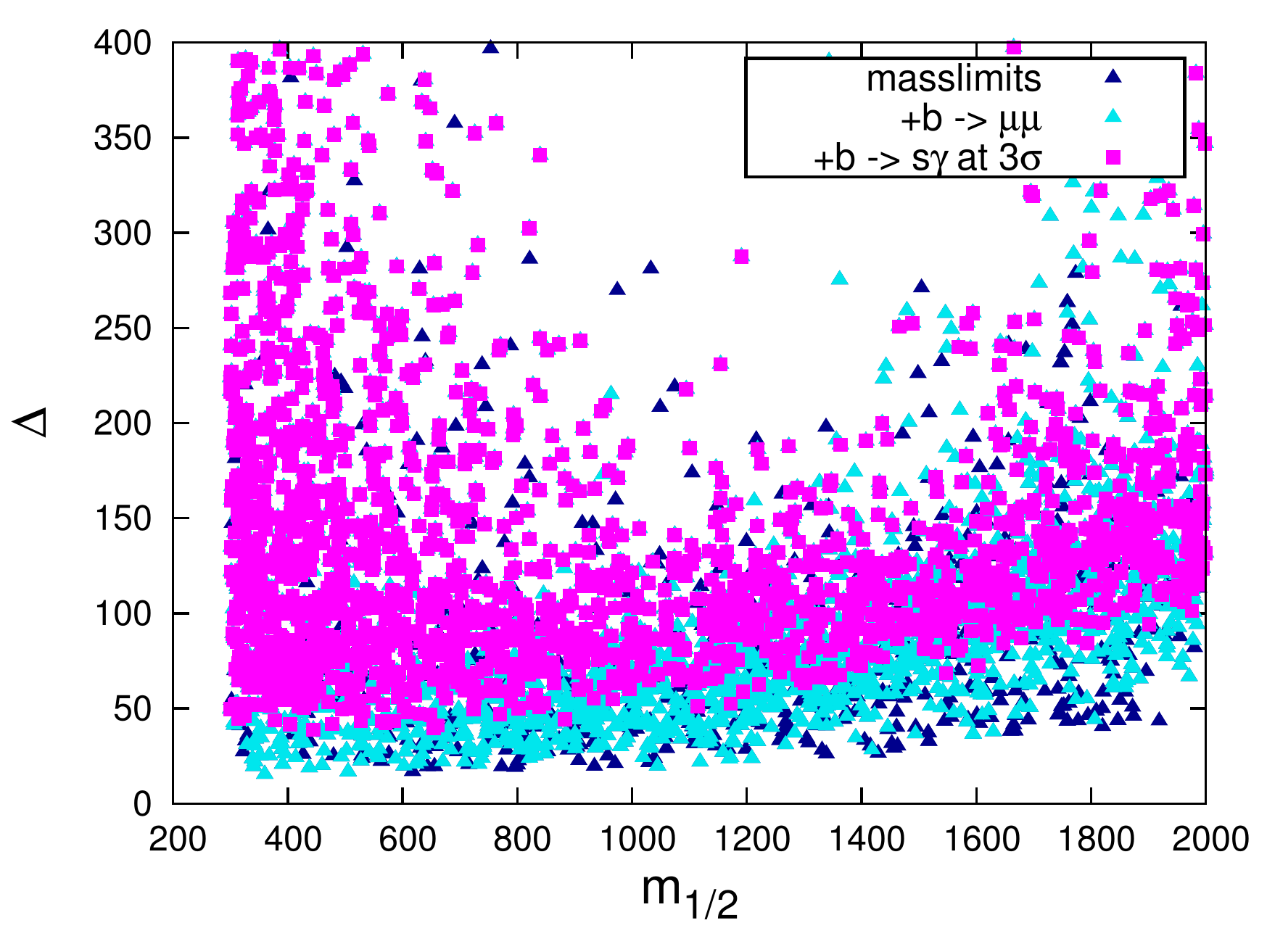}  
  \includegraphics[width=0.37\textwidth,angle=-90]{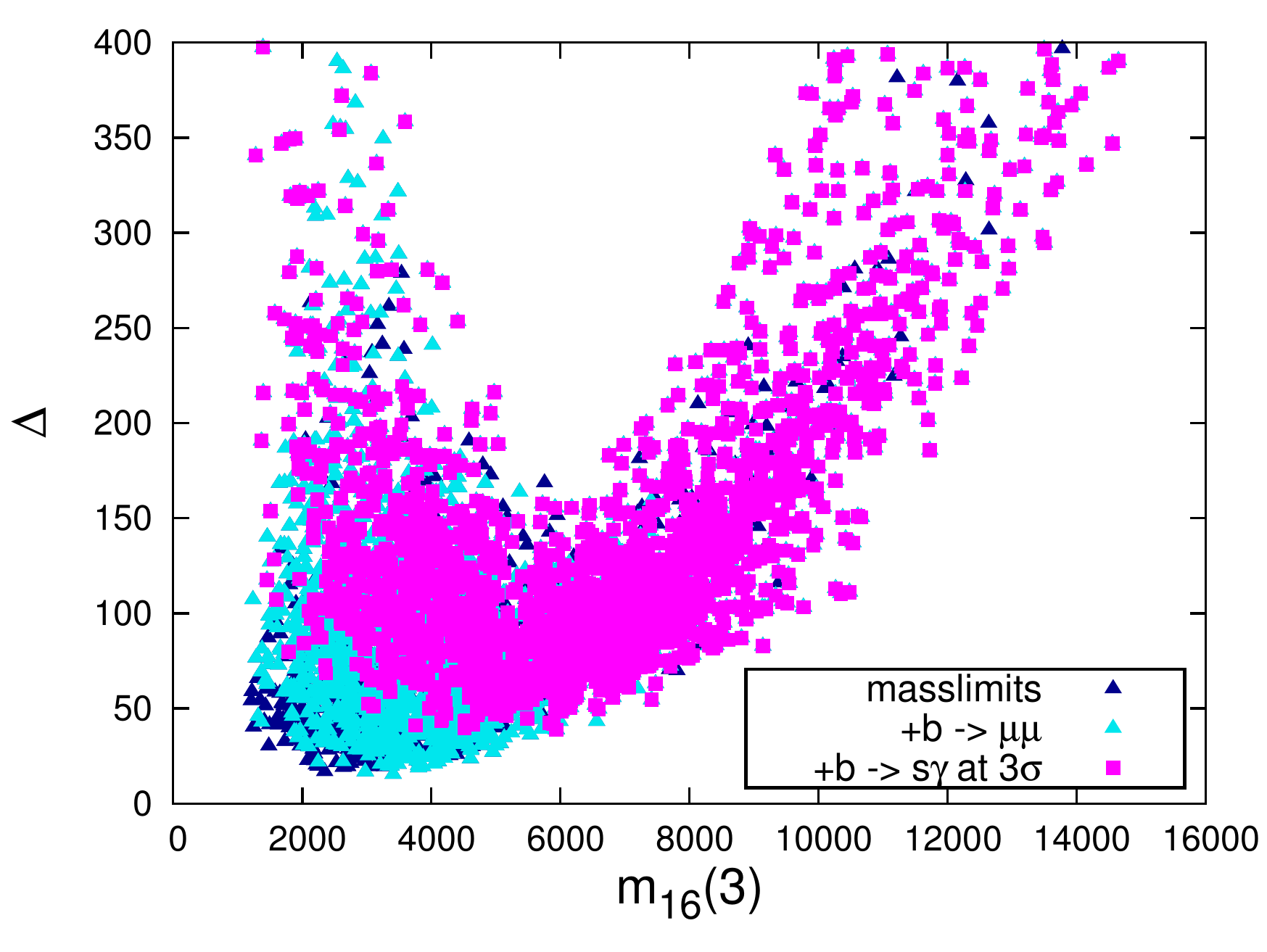}  
  \includegraphics[width=0.37\textwidth,angle=-90]{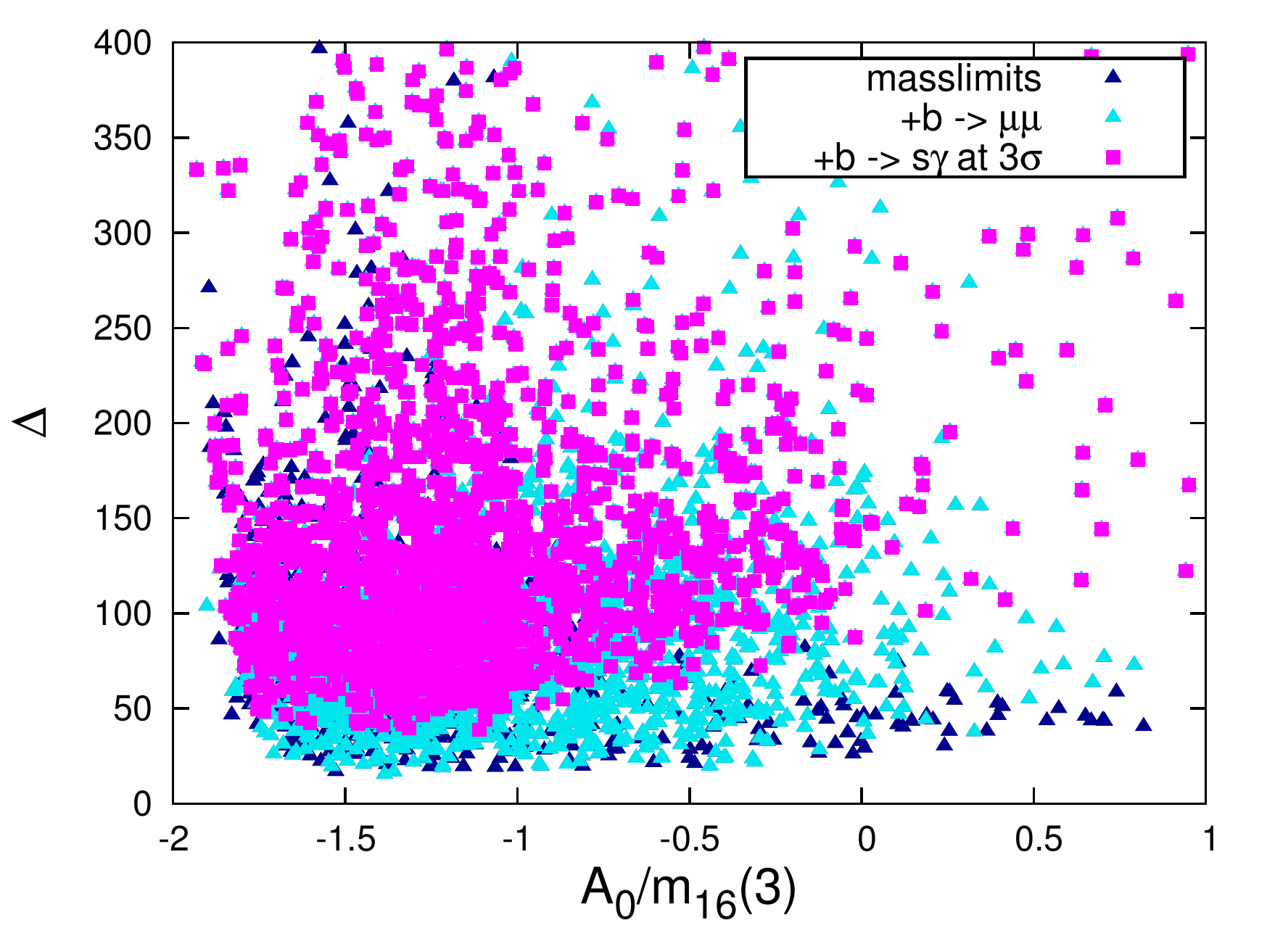}  
  \includegraphics[width=0.37\textwidth,angle=-90]{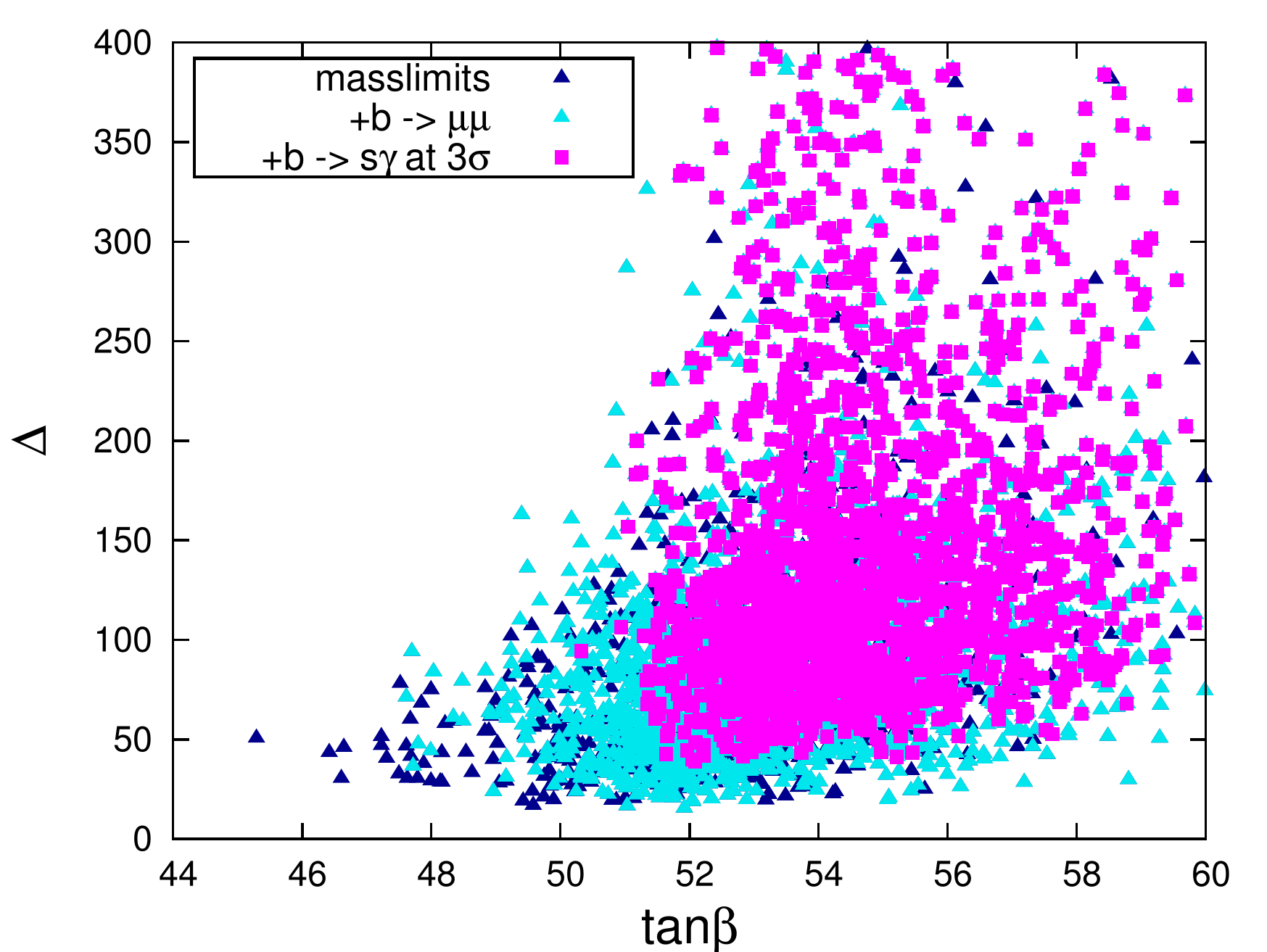}  
\caption{Dependence of fine-tuning $\Delta$ on $m_{1/2}$, $m_{16}(3)$, $A_0/m_{16}(3)$ and $\tan\beta$, for $\Ryuk<1.4$. Same color code as in Fig.~\ref{fig:RyukDelta}. \label{fig:DeltaInpar} }
\end{figure}

The dependence of $\Delta$ on the input parameters is shown in Fig.~\ref{fig:DeltaInpar} for 
points with $\Ryuk<1.4$. In frame {\it a}), we see a mild preference by $\Delta$ for low $m_{1/2}$, 
{\it i.e.}\ the gluino mass can't be too heavy, lest it pushes the stop masses too high, leading to large
$\Sigma_u$. From frame {\it b}), we see low $\Delta$ prefers $m_{16}(3)\sim 2-4$ TeV. If $m_{16}(3)$ is much higher,
then third generation squarks are too heavy to give low EWFT, while if $m_{16}(3)$ is too ight, we generate
tachyonic spectra: the optimal corresponds to $m_{16}(3)\sim 2-4$ TeV. 
For $m_{16}(3)\sim 1-2$ TeV, then sub-TeV top squarks are generated leading to violation of 
$B$-constraints.
In frame {\it c}), we see that low $\Delta$ allows a wide range of $A_0$ unless $B$-constraints are
imposed, in which case $A_0\sim (-2\ {\rm to}\ 0)\times m_{16}(3)$ is again preferred. 
The large negative $A_0$ values lead to larger $m_h$ values and also can lower the EWFT~\cite{Wymant:2012zp}. In frame {\it d}), we see that there is only 
mild preference for $\tan\beta\sim 48-58$ values unless $B$-constraints are respected, in which case
$\tan\beta\agt 52$ is preferred.

\begin{figure}[t]\centering
  \includegraphics[width=0.37\textwidth,angle=-90]{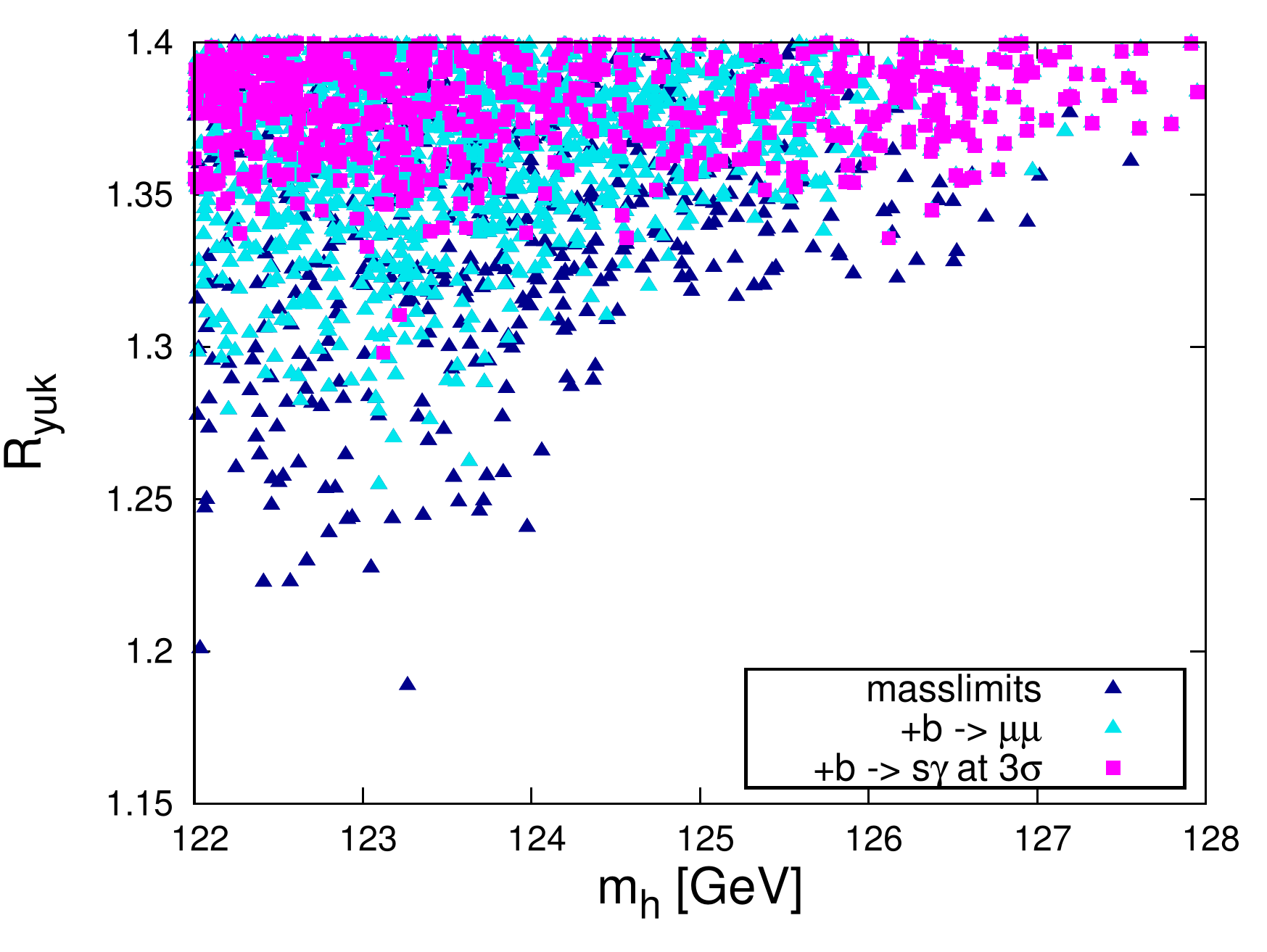}  
  \includegraphics[width=0.37\textwidth,angle=-90]{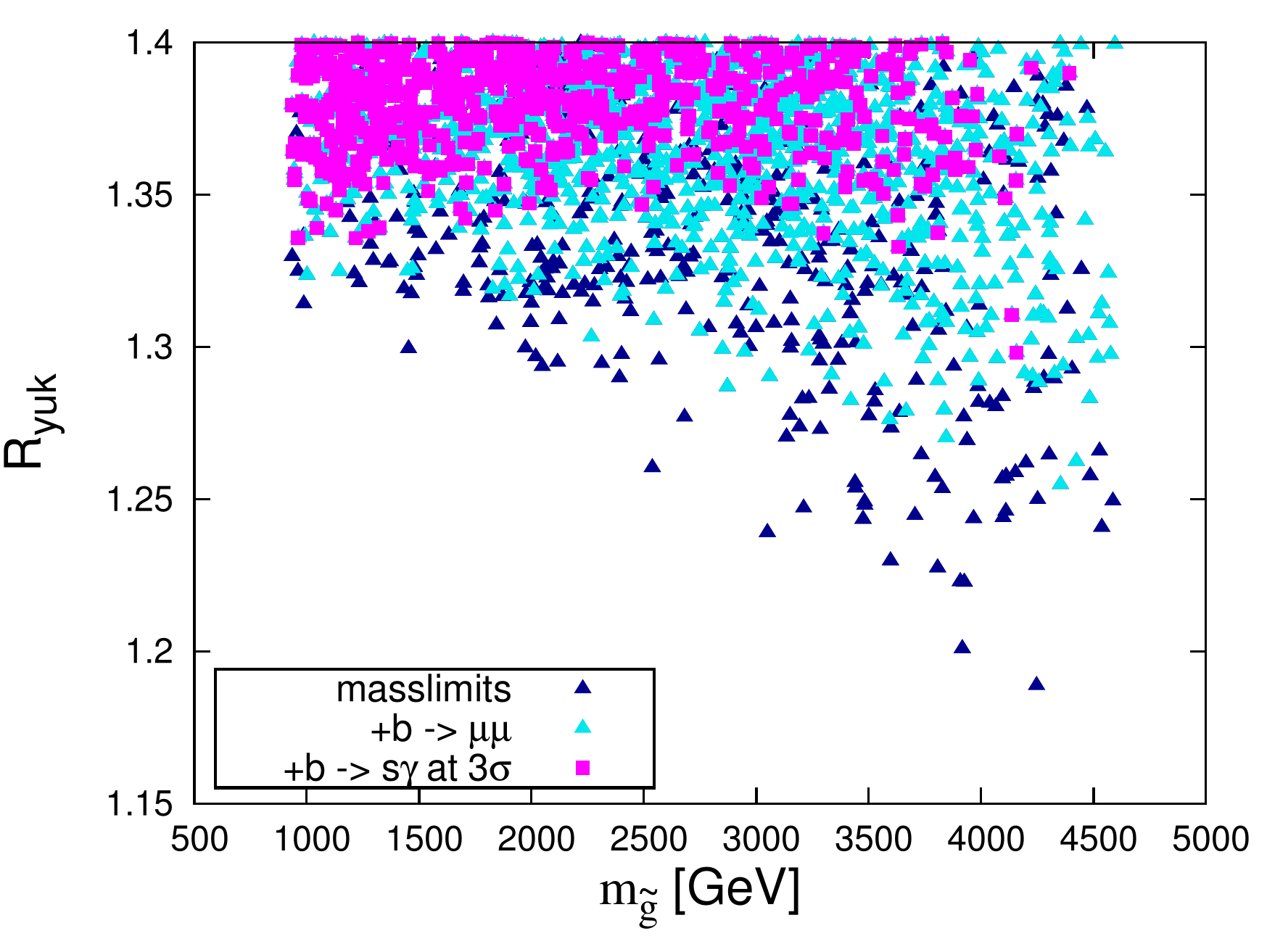}  
  \includegraphics[width=0.37\textwidth,angle=-90]{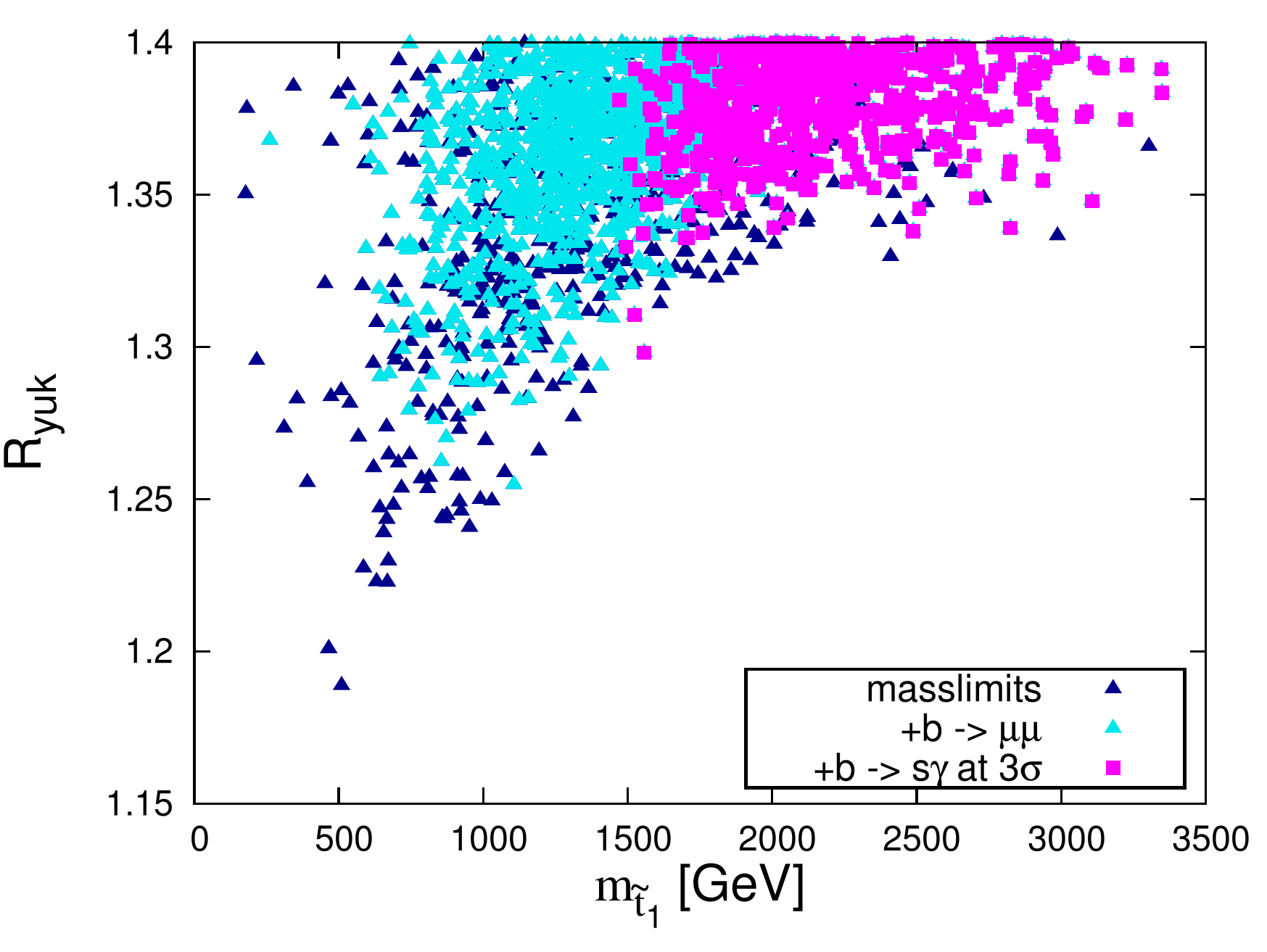}  
  \includegraphics[width=0.37\textwidth,angle=-90]{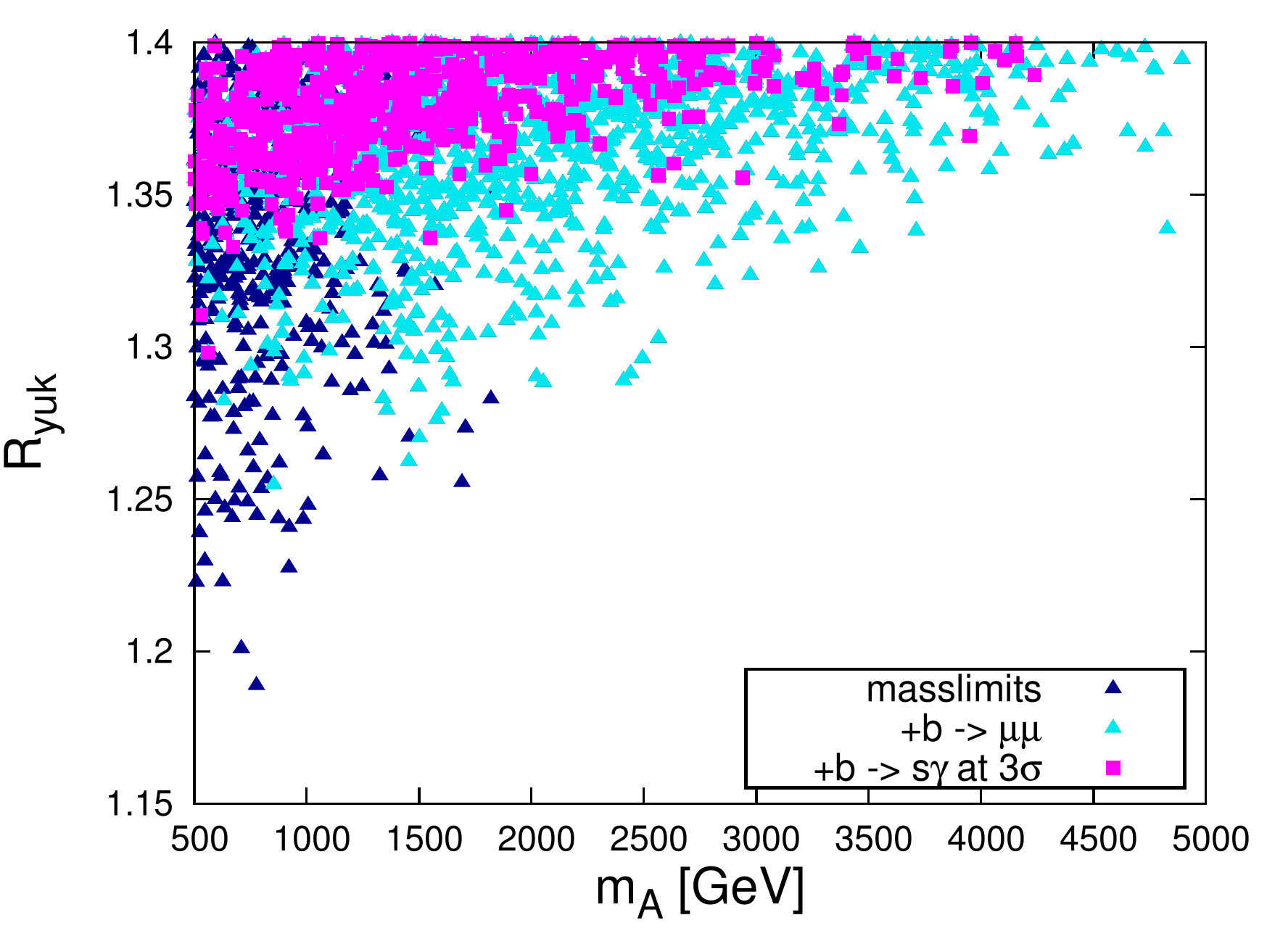}  
\caption{Dependence of $\Ryuk$ on $m_h$, $m_{\tilde g}$, $m_{\tilde t_1}$ and $m_A$,  for $\Delta<100$. 
Same color code as in Fig.~\ref{fig:RyukDelta}. \label{fig:RyukMasses} }
\end{figure}

In Fig. \ref{fig:RyukMasses} we show $\Ryuk$ versus various sparticle and Higgs masses.
In frame {\it a}), the distribution versus $m_h$ is shown, and we see that low $\Ryuk$ prefers
the lower range of $m_h$. This is understandable since low EWFT prefers lower top and bottom squark
masses, which may not feed a sufficient radiative correction into the $m_h$ computation. These low 
third generation squark mass solutions also tend to give large contributions to $B$-constraint. If we 
impose $B$-constraints, then $m_h$ can live in the $122-128$ GeV range, at the cost of larger $\Ryuk$.
In frame {\it b}), we show the distribution versus $m_{\tg}$. While all solutions---especially low $\Ryuk$ ones---favor the heavier range of $m_{\tg}$ ($m_{\tg}\sim 2-5$~TeV, likely beyond LHC reach), the solutions obeying $B$-constraints
tend to slightly favor lower $m_{\tg}$, possibly within range of LHC14 searches.
In frame {\it c}), the distribution versus $m_{\tst_1}$ is shown. Here, we see a clear demarcation:
$B$-constraints favor $m_{\tst_1}\agt 1.5$ TeV to suppress SUSY loop contributions to $\bsg$. 
This constraint forces the minimum $\Ryuk$ to move form $\sim 1.18$ to about $1.3$. 
Additional small flavor-violating contributions to the MSSM Lagrangian could alter the predicted 
$\bsg$ and/or $\bmm$ rates and thus allow the lower $\Ryuk$ solutions\cite{Gabbiani:1996hi}. 
In frame {\it d}), we show the distribution versus pseudoscalar.
Higgs mass $m_A$. We see low $\Ryuk$ favors the lower range of $m_A$, although values up to and beyond
5~TeV are also possible (at the cost of $\Ryuk\sim 1.4$, however). 
Since $\tan\beta\sim 50$, LHC searches for $A,\ H\to\tau^+\tau^-$ will access a significant range of $m_A$ in this case.  
It is also possible for some range of $m_A\lesssim 1$~TeV for LHC to 
access $bA\to b\mu^+\mu^-$ production~\cite{Baer:2011ua}.

\begin{figure}[t]\centering
  \includegraphics[width=0.37\textwidth,angle=-90]{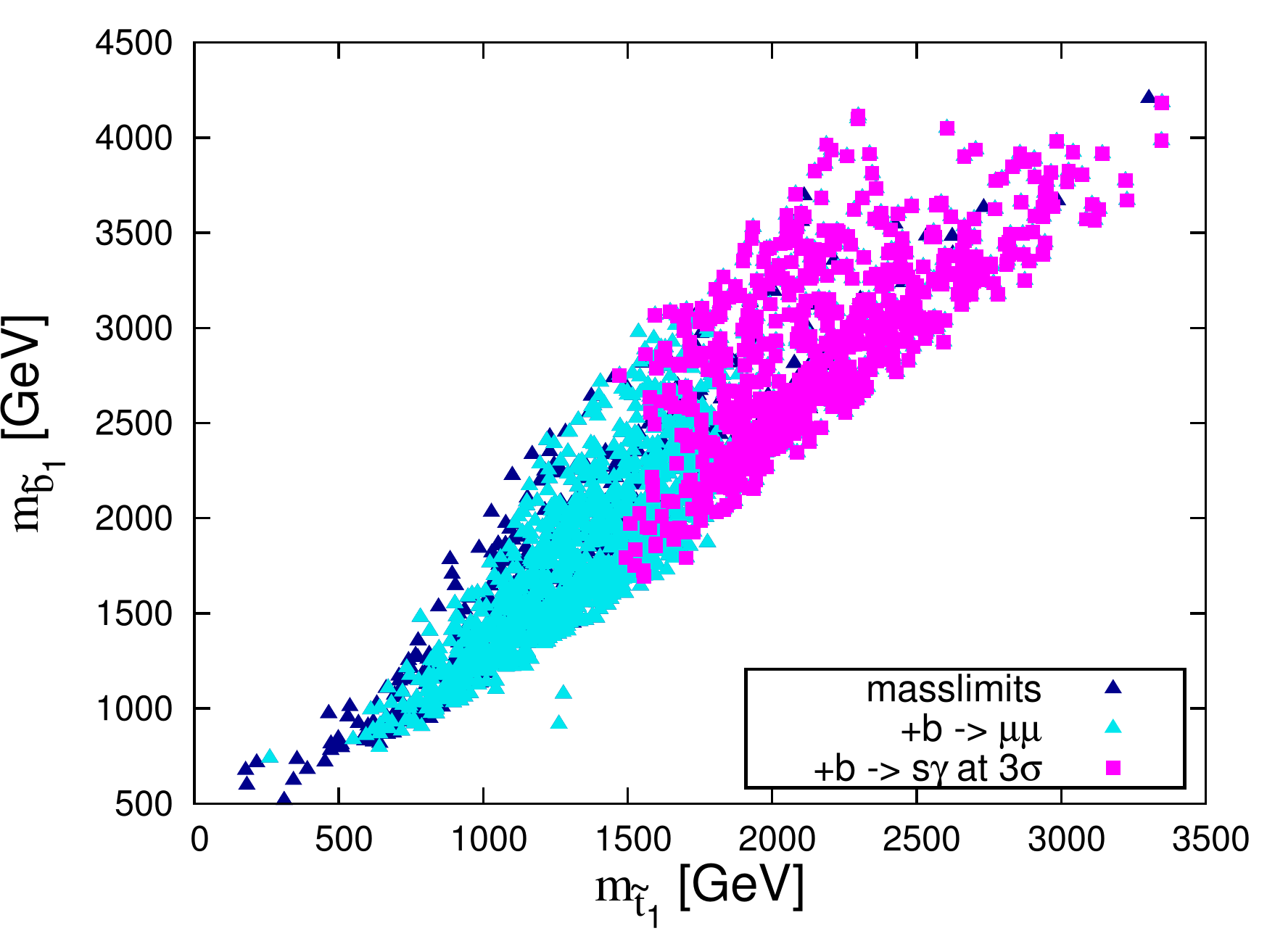}  
  \includegraphics[width=0.37\textwidth,angle=-90]{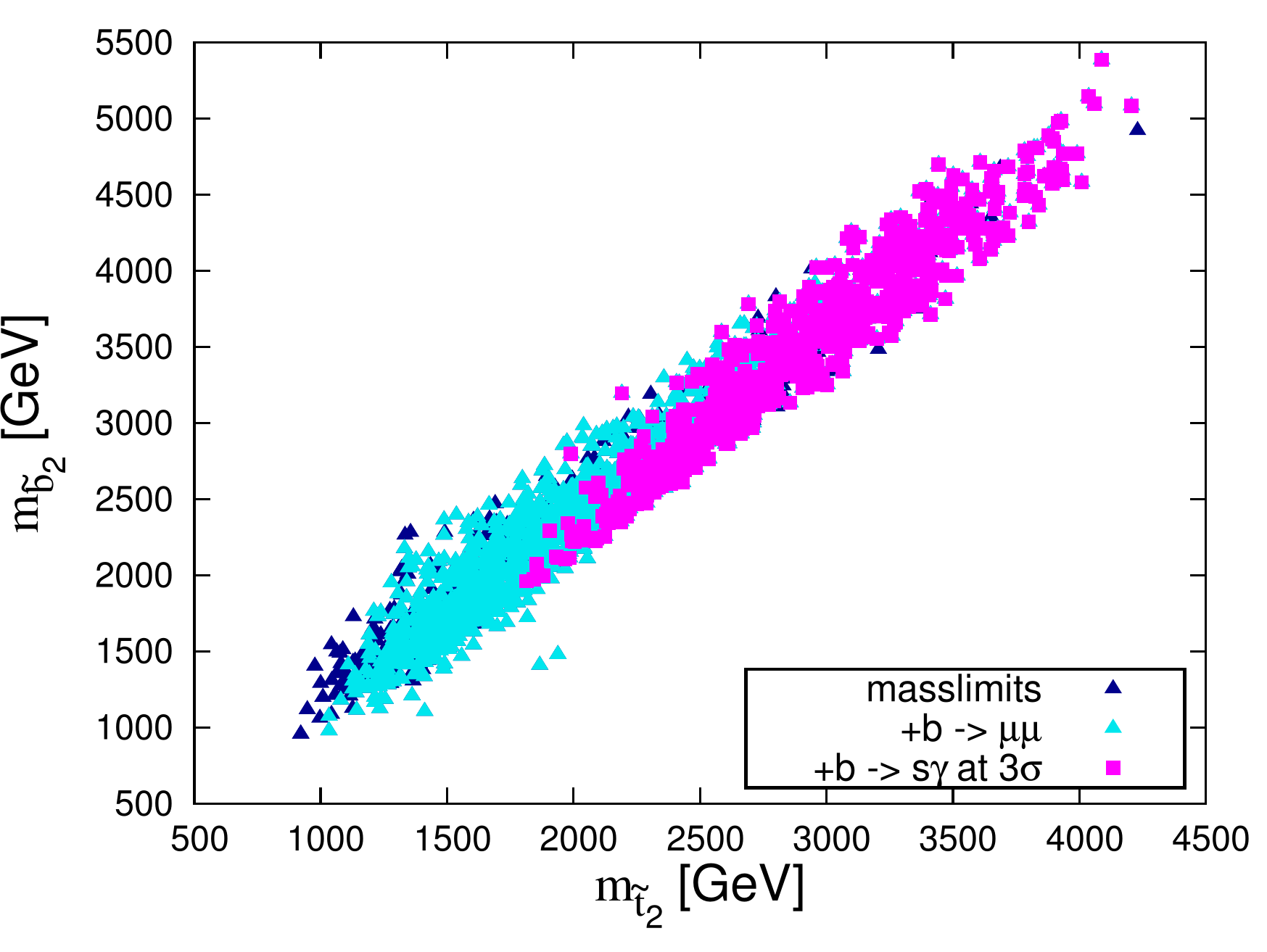}  
\caption{Distribution of scan points in {\it a}) $m_{\tst_1}\ vs.\ m_{\tb_1}$ space and
{\it b}) $m_{\tst_2}\ vs.\ m_{\tb_2}$ space for $\Ryuk<1.4$ and $\Delta <100$. 
Same color code as in Fig.~\ref{fig:RyukDelta}. \label{fig:t1b1} }
\end{figure}

In Fig. \ref{fig:t1b1}, we show scatter plots of YUNS points with $\Ryuk<1.4$ and $\Delta <100$ in
{\it a}) $m_{\tst_1}\ vs.\ m_{\tb_1}$ and {\it b}) $m_{\tst_1}\ vs.\ m_{\tb_1}$ space.
From frame {\it a}), we see as expected that $m_{\tst_1}$ and $m_{\tb_1}$ are correlated, 
with some solutions reaching well below $m_{\tst_1}\sim 500$ GeV, which should be accessible to LHC searches.
However, in this case these points all violate $B$-constraints, so that requiring $B$-constraints within
measured range requires instead $m_{\tst_1},\ m_{\tb_1}\agt 1.5$ TeV, likely beyond the 14~TeV LHC reach.
In frame {\it b}), we see also that $m_{\tb_2}$ and $m_{\tst_1}$ are correlated. This is different from
usual NS, where $m_{\tb_2}$ can be far above $m_{\tst_{1,2}}$ and $m_{\tb_1}$. The reason here is that
$f_b$ is large and so there is also a non-negligible contribution to $\Sigma_u$ from $\tb_{1,2}$.
Imposing $B$-constraints, we find $m_{\tst_2}$ and $m_{\tb_2}$ both $\agt 2$ TeV (the former aids in 
lifting $m_h$ into its measured range). 

\section{Benchmark points}
\label{sec:bm}

In this section we present several YUNS benchmark points, see Table~\ref{tab:bm}, 
and compare to one YUS benchmark point (HSb, from the ``just-so'' HS model) 
from Ref.~\cite{Baer:2008yd}. 

For HSb, $\Ryuk \sim 1.02$, which is nearly perfect Yukawa coupling unification.
The Higgs mass $m_h\simeq 127.8$ GeV is also sufficiently heavy. Unfortunately, the point is now
excluded by LHC7 searches for multi-jet$+\eslt$ plus one $b$-tag searches, since $m_{\tg}$ is only
351~GeV. We also list in Table~\ref{tab:bm} the EWFT measure; for HSb we have $\Delta =2489$,
indicating exceptionally high level of fine-tuning. Much of this comes from the $\mu$ parameter which
turns out to be of order 3~TeV, and thus requires a large value of $m_{H_u}^2$ at the weak scale to cancel against.
                                                                             
\begin{table}\centering
\begin{tabular}{lcccc}
\hline
parameter & HSb & YUNS1 & YUNS2 & YUNS3\\
\hline
$m_{16}(1,2)$& 10000 & 19390.0 & 17149.4 & 19928.8\\
$m_{16}(3)$  & 10000 & 5938.8 & 2490.4 & 2490.5 \\
$m_{1/2}$  & 43.9 & 444.4 & 1859.2 & 1809.1 \\
$A_0$      & $-19947.3$ & $-6595.5$ & $-1374.1$ & $-319.1$ \\
$\tan\beta$& 50.398 & 52.09 & 54.2& 51.7 \\
$\mu$      & 3132.6 & 136.0 & 106.1 & 169.6 \\
$m_A$      & 1825.9 &  884.1 & 541.2 & 776.4\\
\hline
$f_t$      & 0.557 & 0.578 & 0.563 & 0.549 \\                                                
$f_b$      & 0.557 & 0.423 & 0.474 & 0.496 \\                                                
$f_\tau$   & 0.571 & 0.561 & 0.618 & 0.590 \\                                                
$\Ryuk$        & 1.025  & 1.37 & 1.3 & 1.19 \\
$\Delta$   & 2489 & 39 & 105 & 48 \\
\hline
$m_{\tg}$   & 351.2 & 1328.5 & 4309.5 & 4247.2 \\
$m_{\tu_L}$ & 9972.1 & 19396.8 & 17466.6 & 20189.0 \\
$m_{\tst_1}$& 2756.5 & 1587.3 & 1598.0  & 509.3\\
$m_{\tb_1}$ & 3377.1 & 2176.3 & 1857.1 & 793.1\\
$m_{\te_R}$ & 10094.7 & 19379.3 & 17153.2 & 19930.2\\
$m_{\tw_1}$ & 116.4 & 137.3 & 112.5 & 177.4\\
$m_{\tz_2}$ & 113.8 & 147.4 & 112.0 & 176.2 \\
$m_{\tz_1}$ & 49.2 &  118.9  & 106.2 & 170.3\\
$m_h$       & 127.8 &  123.1 & 123.8 & 123.3 \\
\hline
$\bsg$ & $3.1\times 10^{-4}$ & $2.7\times 10^{-4}$ & $2.8\times 10^{-4}$& $2.4\times 10^{-4}$ \\
$\bmm$ & $8.1\times 10^{-9}$ & $4.5\times 10^{-9}$ & $4.6\times 10^{-9}$& $2.4\times 10^{-8}$ \\
$\Omega_{\tz_1}^{\rm TP}h^2$ & 4613 & 0.01 & 0.004 & 0.007 \\
$\sigma^{\rm SI}(\tz_1p)$ pb & $2.2\times 10^{-13}$ & $4.5\times 10^{-8}$& $4.7\times 10^{-9}$ & $2.9\times 10^{-9}$\\
$\sigma^{\rm SD}(\tz_1p)$ pb & $1.2\times 10^{-9}$& $7.3\times 10^{-4}$& $3.6\times 10^{-5}$ & $1.6\times 10^{-5}$\\
$\langle\sigma v\rangle |_{v\to 0}$ ${\rm cm^3/s}$ & $1.9\times 10^{-32}$ & $2.4\times 10^{-25}$& $3.3\times 10^{-25}$ 
& $2.8\times 10^{-25}$\\
\hline
\end{tabular}
\caption{Parameters and masses in~GeV units
for HSb~\cite{Baer:2008yd} and three Yukawa-unified natural SUSY (YUNS) benchmark points.
We also show $B$-decay constraints and dark matter relic density and (in)direct detection cross sections.
\label{tab:bm} }
\end{table}

In contrast, point YUNS1 in column~3 has low fine-tuning of $\Delta =39$, at the cost of relaxing $\Ryuk$ to $1.37$.
For YUNS1, $m_{\tg}\simeq 1.3$ TeV, with first/secnd generation squarks at $\sim 19$ TeV. The point is
likely beyond LHC8 reach, but should be accessible to LHC14 with 10--100~fb$^{-1}$. 
In column~4, we list YUNS2 with $\Ryuk =1.3,$ as low as allowed by $B$-constraints, but with
$\Delta\sim 100$. This point has $m_{\tg}\sim 4.3$~TeV and $m_{\tq}\sim 17.5$ TeV, so it is likely beyond
LHC reach, including a high-luminosity upgrade. While the higgsino-like chargino is only $112.5$~GeV, it decays via 3-body mode into a higgsino-like $\tz_1$ with $m_{\tz_1}=106.2$ GeV, so that visible decay products are very soft,
and likely impossible to observe above SM backgrounds at the LHC.
Point YUNS3, listed in column~5, features $\Ryuk$ as low as 1.19, with $\Delta =48$.
This point has $\bsg=2.4\times 10^{-4}$, and $\bmm =2.4\times 10^{-8}$, so it falls  
out of the $B$-physics allowed range. While gluinos and first/second generation squarks are beyond
LHC reach, the rather light top and bottom squarks may be accessible to LHC searches.
All these points have $\Omega_{\tz_1}h^2\ll0.11$, leaving room for non-thermal higgsino
production and axions. This contrasts point HSb, which has a much too thermal abundance and so would need an 
extremely light axino or huge late-time entropy production to tame this over-abundance~\cite{Baer:2011uz}.

\begin{figure}[t]\centering
  \includegraphics[width=0.37\textwidth,angle=-90]{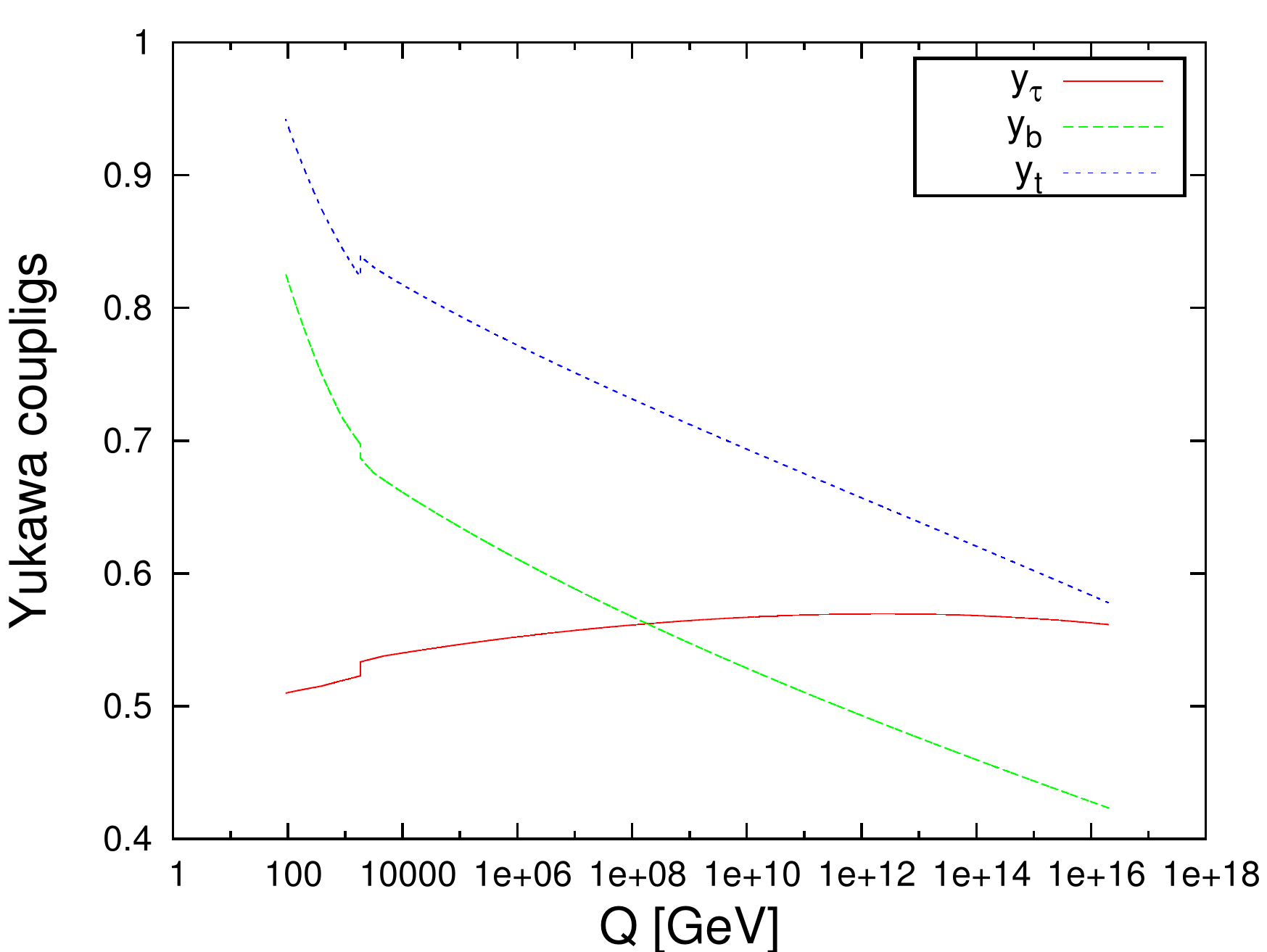}  
  \includegraphics[width=0.37\textwidth,angle=-90]{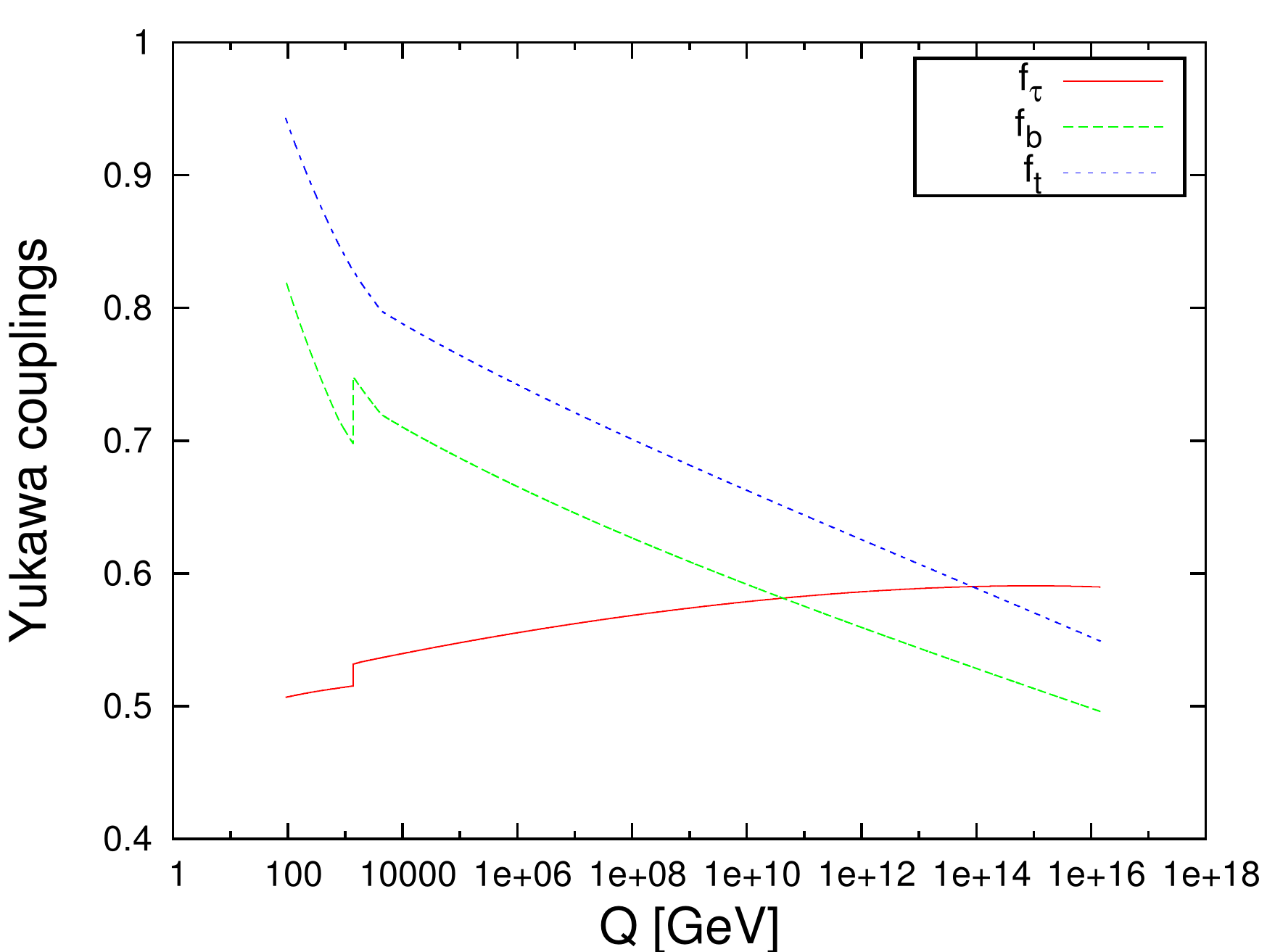}  
\caption{Evolution of Yukawa couplings for benchmark points YUNS1 (left) and YUNS3 (right)
versus renormalization scale $Q$. \label{fig:evol} }
\end{figure}

In Fig.~\ref{fig:evol}, we show the Yukawa coupling evolution of $f_t$, $f_b$ and $f_\tau$ versus
renormalization group scale $Q$, from $m_{\rm weak}$ to $M_{\rm GUT}$ for benchmark  points YUNS1 and YUNS3. 
These can be compared to similar plots for YUS, as in {\it e.g.} Fig.~6 of Ref.~\cite{Baer:2009ie}. 
The SUSY threshold corrections implemented at the 
scale $Q=\sqrt{m_{\tst_1}m_{\tst_2}}$ show up as jumps in the curves. In the case of YUS models,
the $m_b$ threshold correction is positive due to a large $\tst_i\tw_j$ loop contribution which goes like
$\delta_b\sim (f_t^2/32\pi^2)(\mu A_t/m_{\tst}^2)\tan\beta$, where both $\mu$ and $A_t$ are extremely large.
For YUNS models, with rather low $\mu$, these loops are suppressed and in the case of YUNS1, 
the $\tg\tb_i$ loops actually dominate, and are of opposite sign to the $\tst_i\tw_j$ loops, 
leading to the slight downward jump of $f_b$ and thus bad Yukawa coupling unification. 
For YUNS3, the $\tst_i\tw_j$ loops are larger, and the jump goes upwards, thus providing better Yukawa coupling unification. 

\clearpage
\section{Yukawa-unified natural SUSY: LHC, ILC and DM searches}
\label{sec:exp}

In this Section, we discuss the observable consequences of YUNS for LHC, ILC and
direct and indirect WIMP and also axion searches.

\subsection{YUNS at LHC}
\label{ssec:lhc}

In natural SUSY models, it is favorable to have multi-TeV first/second generation
squarks and sleptons because 1.~they are safely beyond current LHC searches, 2.~they
provide at least a partial solution to the SUSY flavor and CP problems and 
3.~they provide additional suppression of third generation scalar masses via large 2-loop RGE effects.
However, this means they are likely beyond any conceivable LHC reach. Third generation squarks may be much
lighter, and in generic NS models are naively expected to be below the TeV scale (but see Ref.~\cite{Wymant:2012zp}
where 1--4~TeV third generation squarks work just fine, and lift the value of $m_h$ into its measured range).
In the case of YUNS, with $\tan\beta\sim 50$, the combined light squarks and large $\tan\beta$ usually
imply violation of $B$-constraints, and if these are imposed, then top and bottom squarks are
beyond 1.5 TeV, and likely inaccessible to LHC searches. If additional sources of flavor violation are invoked,
then the $B$-constraints may be invalid, and then the solutions with much lower $\Ryuk\sim 1.2$ are accessible,
along with much lighter top and bottom squarks, potentially accessible to LHC searches.

For YUNS, the gluino mass may lie anywhere in the 1--5~TeV range. It has been estimated in Ref.~\cite{lhcreach}
that LHC14 with 100 fb$^{-1}$ should be able to access gluino pair production in the case of heavy squarks
for $m_{\tg}$ up to 1.8~TeV. In the case of YUNS with heavier top and bottom squarks, the 
$\tg$ is expected to dominantly decay via 3-body modes into $t\bar{t}\tz_i$ and $tb\tw_i$ final states.
The gluino pair events will thus contain multi-jets plus missing energy plus isolated leptons plus
several identifiable $b$-jets~\cite{Baer:1990sc}.
While the higgsino-like chargino and neutralino production cross sections can be large, 
their decays to soft visible particles, arising from the small energy release in their 3-body decays, will be 
difficult to detect at LHC above SM backgrounds~\cite{bbh}.

In the case where $m_A\alt 1$ TeV, then it may be possible to detect $A,\ H\to\tau^+\tau^-$, especially
if these are produced in association with $b$-jets, {\it e.g.} $pp\to bA,\ bH$ production.
It may also be possible to detect $bA,\ bH$ production with $A,\,H\to\mu^+\mu^-$~\cite{Baer:2011ua}, 
since the production and decay
are enhanced at large $\tan\beta$. In this case, the $A,\ H$ mass and width may be determined by 
reconstructing $m(\mu^+\mu^- )$. At large $\tan\beta$, this width is typically in the tends of GeV range
and is very sensitive to $\tan\beta$. This reaction offers a method to distinguish YUNS from NS, in that
the former is expected to occur at $\tan\beta\sim 50$.

\subsection{YUNS at ILC}
\label{ssec:ilc}

A linear $e^+e^-$ collider operating at $\sqrt{s}\sim 250-500$ GeV would
in many ways be an optimal discovery machine for YUNS. The reason is that by construction
$\mu\alt 250$ GeV, so chargino and neutralino pair production should always be available.
While the small energy release in $\tw_1$ and $\tz_2$ decay is problematic at LHC, it should be
much more easily observable in the clean environment of an $e^+e^-$ collider. In this sense, an
$e^+e^-$ collider operating at $\sqrt{s}\sim 250-500$ GeV would be a higgsino in addition to a
Higgs factory. It is also possible that some lighter third generation squarks are accessible
to ILC with $\sqrt{s}\sim 1$ TeV or CLIC with $\sqrt{s}=3$ TeV, 
depending if one avoids $B$-constraints and accepts the low mass, low $\Ryuk$ solutions.

\subsection{Higgsino-like WIMPs}
\label{ssec:dm}

\begin{figure}[t]\centering
  \includegraphics[width=0.37\textwidth,angle=-90]{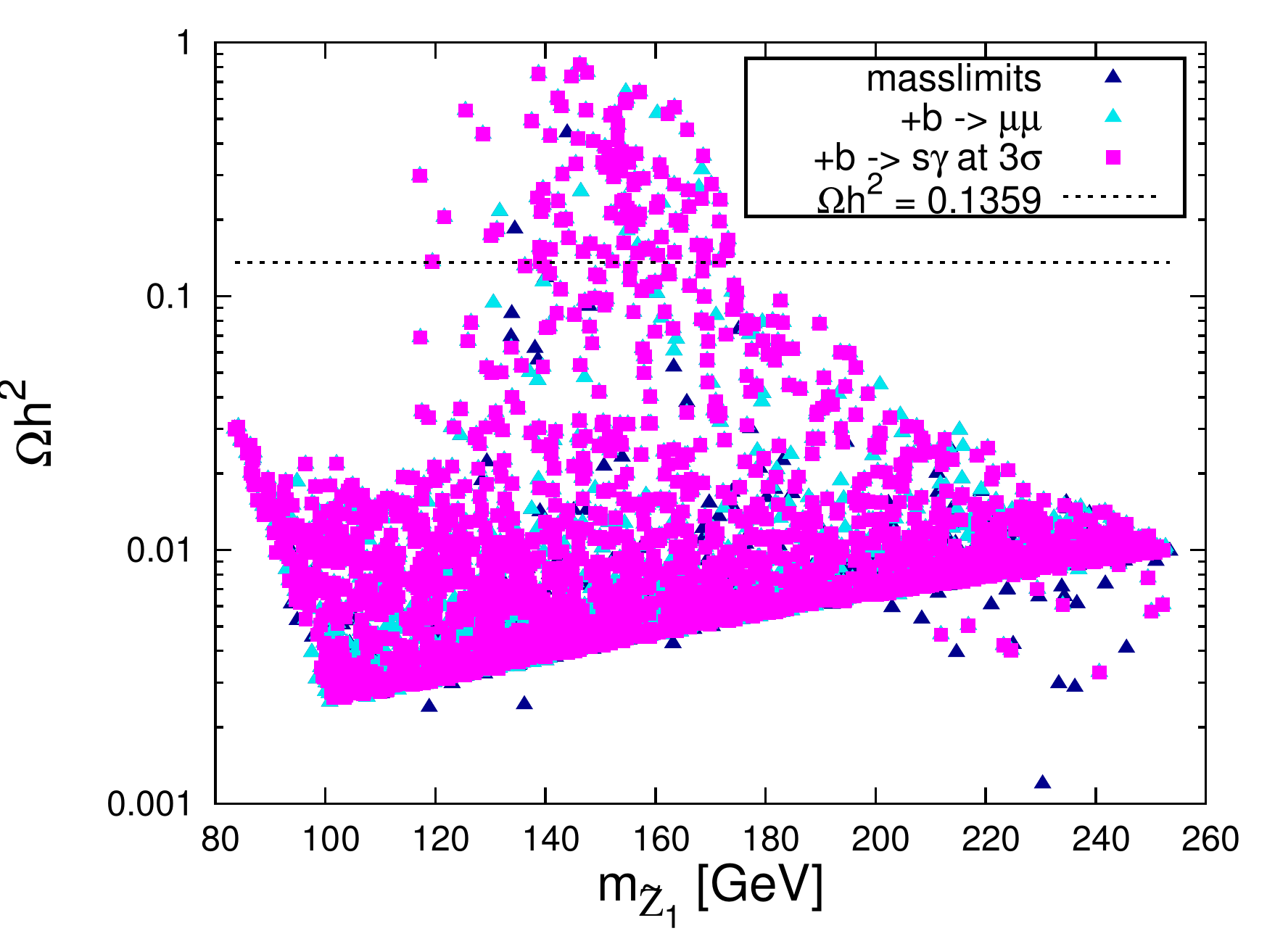}  
  \includegraphics[width=0.37\textwidth,angle=-90]{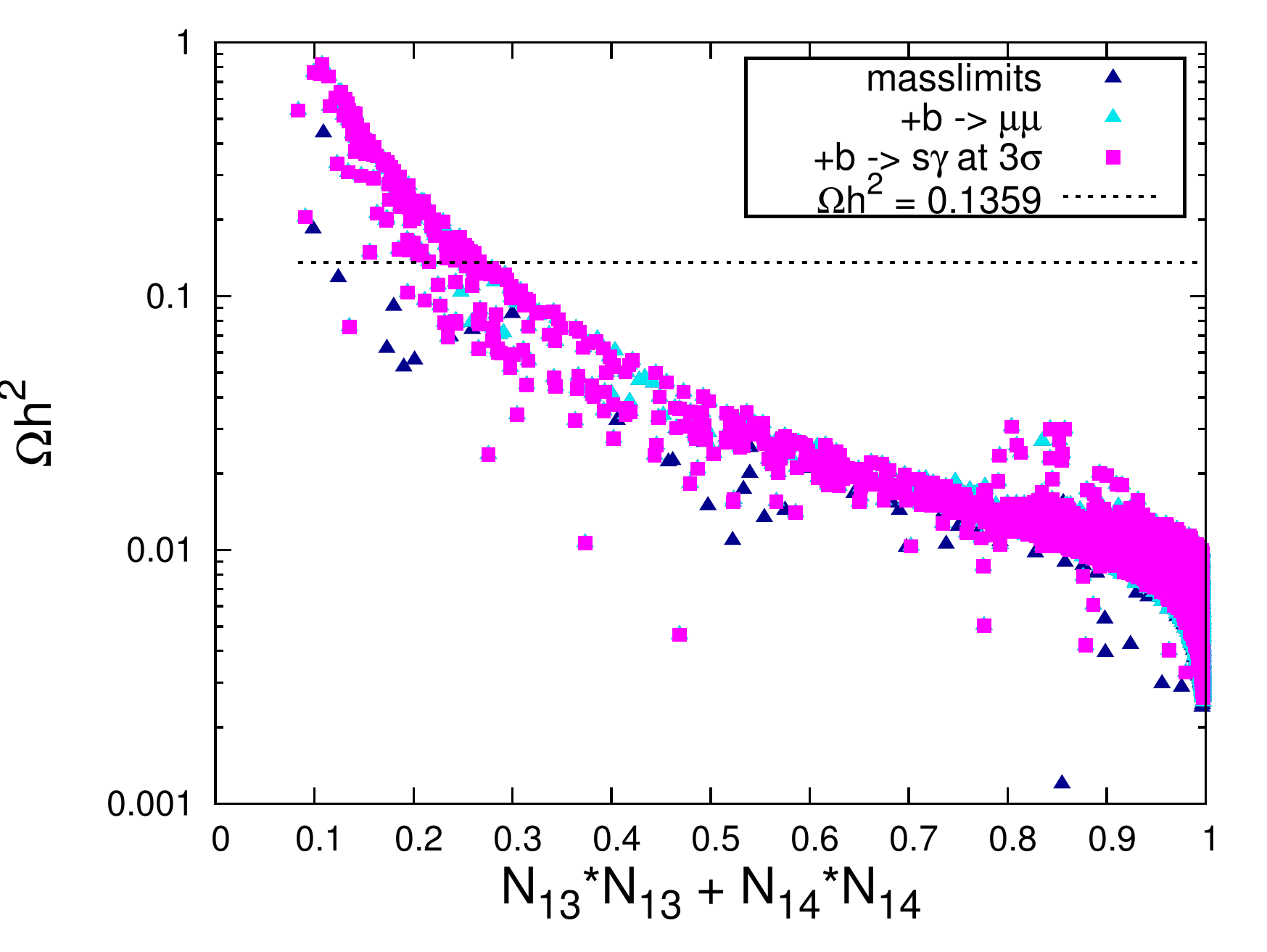}  
\caption{Thermal neutralino relic density $\Omega h^2$ versus $\tz_1$ mass (left) and versus $\tz_1$ higgsino fraction (right) for scan points with $\Ryuk<1.4$ and $\Delta <100$. 
Same color code as in Fig.~\ref{fig:RyukDelta}.
}
\label{fig:relic} 
\end{figure}

A generic prediction of both NS and YUNS models is that the LSP is a higgsino-like WIMP
with a typical under-abundance of thermally produced (TP) neutralinos 
$\Omega_{\tz_1}^{\rm TP}h^2\sim 0.002-0.01$. 
However, in cases where $m_{1/2}$ is
as low as $\sim 300$ GeV and $\mu$ is as large as $200-250$ GeV, 
then there can be substantial bino--higgsino mixing, boosting $\Omega_{\tz_1}^{\rm TP}h^2$ up to $0.11$ or even beyond.
The situation is illustrated in Fig.~\ref{fig:relic}, where we plot $\Omega_{\tz_1}^{\rm TP}h^2$ versus $m_{\tz_1}$ 
and versus the $\tz_1$ higgsino fraction 
from YUNS models satisfying $\Ryuk<1.4$ and $\Delta <100$.

The typical under-abundance is an appealing feature if one invokes the Peccei-Quinn solution to the
strong CP problem, in which case one must introduce an axion superfield $\hat{a}$ which contains a
pseudoscalar axion $a$ as well as a spin-1/2 axino $\ta$ and a spin-0 saxion $s$.
In such models, one expects both saxion and axino masses at or around the SUSY breaking scale, so that dark matter
is comprised of an axion-WIMP admixture. In this case, thermal production of axinos and subsequent decay to
states such a $\tg g$ bolster the WIMP abundance beyond its TP-value. In addition, axions are produced
as usual via coherent oscillations. It is also possible to suppress the WIMP abundance in the mixed
$a\tz_1$ cosmology via late-time entropy production from saxion production and decay, although this
case tends to be highly constrained by maintaining successful Big Bang nucleosynthesis (BBN).
The upshot is that in the mixed $a\tz_1$ dark matter scenario, it may be possible to detect both
a WIMP and an axion.

In the case of the YUNS model, the higgsino-like neutralinos have a substantial spin-independent (SI)
direct detection cross section, as illustrated in Table \ref{tab:bm}, where
$\sigma^{\rm SI} (\tz_1 p)\sim 10^{-8}$ pb. While this level of direct detection cross section is now
highly constrained by recent XENON100~\cite{Xenon:2012nq} results, one must bear in mind 
that the higgsino-like WIMPs
would constitute only a portion of the dark matter, so their local abundance might be
up to a factor of 30 lower than is commonly assumed.
In Table \ref{tab:bm}, we also list $\sigma^{\rm SD}(\tz_1 p)$ (relevant for WIMP detection at IceCube)
and $\langle \sigma v\rangle |_{v\to 0}$, relevant for detection of dark matter annihilation into
gamma rays or anti-matter throughout the cosmos. While these cross sections are also at potentially
observable levels, again one must take into account that the overall WIMP abundance may be up 
to a factor of about 30 below what is commonly assumed. 
Plots of $\sigma^{\rm SI} (\tz_1 p)$, $\sigma^{\rm SD}(\tz_1 p)$ and
$\langle \sigma v\rangle |_{v\to 0}$ have been presented in the case of higgsino-like WIMPs in
Ref's~\cite{bbh} and \cite{Baer:2012uy} and so similar plots will not be reproduced here.

\section{Summary and conclusions}
\label{sec:conclude}

Previous analyses of $t-b-\tau$ Yukawa-unified models suffer from two problems: 1. they tend to predict
a light gluino $m_{\tg}\alt 500$ GeV (alhough solutions are possible for much heavier gluinos) which
is now excluded by LHC searches, and 2. they suffer from extreme fine-tuning in the electroweak sector.
In this paper, we examined how well the Yukawa couplings could unify in the natural SUSY context, 
where $\mu\sim 100-250$ GeV, while at the same time requiring the light Higgs mass $m_h\sim 122-128$ GeV. 
The small value of $\mu$ suppresses the large $\tst_i\tw_j$ loop contributions
to the $b$-quark Yukawa coupling which seem to be needed for precision Yukawa coupling unification. 
Nonetheless, by scanning over NUHM2 parameters with split third
generation, we are able to find solutions with $\Ryuk$ as low as 1.18. These solutions, with very light
third generations squarks and $\tan\beta\sim 50$ tend to violate $B$-physics constraints. If $B$-physics
constraints are imposed, then only $R\sim 1.3$ can be achieved. However, the $B$-physics calculations
can be modified if additional small flavor-violating terms are allowed in the MSSM Lagrangian, so it is
not clear how seriously $R\gtrsim 1.3$ should be taken.

The Yukawa-unified natural SUSY spectra have important differences from previous YU spectra.
The gluino mass can easily be in the 1--4 TeV range, thus avoiding LHC constraints from SUSY searches.
The light higgsino-like charginos and neutralinos decay to soft particles, also avoiding LHC searches.
However, light higgsinos should be easily accessible to an ILC with $\sqrt{s}\sim 0.25-05$ TeV, 
as is typical of all NS models. In addition, as in all NS models, the lightest neutralino is 
higgsino-like with a typical thermal underabundance of WIMP dark matter. We regard this as a positive feature
in that the WIMP abundance is typically increased in non-standard (but more attractive) cosmologies
such as those conatining mixed axion-neutralino cold dark matter.

The question arises as to how to distinguish YUNS from ordinary NS. 
The YUNS model requires $\tan\beta\sim 50$, which leads to large production cross sections
for heavy Higgs bosons $A$ and $H$ at LHC, and large widths for these particles.
If the rare decays $A,\, H\to\mu^+\mu^-$ can be identified with suficiently high statistics 
(perhaps at a luminosity upgraded LHC), then the widths may be measured
with precision, allowing one to highly constrain $\tan\beta$, and perhaps verify that it is
consistent with YUNS models.

\section*{Acknowledgments}

This work has been supported in part by the Office of Science, US Department of Energy  
and by  IN2P3 under contract PICS FR--USA No.~5872. 
HB and SK acknowledge the hospitality of the Aspen Center for Physics which is supported by the National Science Foundation Grant No. PHY-1066293.


\end{document}